\documentclass[twocolymn]{aa} % for a normal version

\pdfoutput=1

\usepackage{graphicx}
\usepackage{epstopdf}
\usepackage{txfonts}
\usepackage{natbib}
\bibpunct{(}{)}{;}{a}{}{,} % to follow the A&A style
\usepackage[normalem]{ulem} % For editing purposes
\newcommand{\event} {MOA-2007-BLG-192}
\newcommand{\planet} {MOA-2007-BLG-192Lb}
\newcommand{\host} {MOA-2007-BLG-192L}
\newcommand{\planetjulia} {MOA-2008-BLG-310Lb}

\def\PRE  {\tilde{r}_E}

\def\msun      {M_{\odot}}
\def\mTer      {M_{\oplus}}
\def\mearth      {M_{\oplus}}

\def\change    { }
\begin{document}
 \title{A frozen super-Earth orbiting a star at the bottom of the Main Sequence\thanks{Based on observations under ESO Prog.IDs: 279.C-5044(A) and 383-C.0495(A)}.
 }
 \author{D. ~Kubas \inst{1,2}, J. ~P. Beaulieu \inst{1,3}, D.~P.~Bennett\inst{4}, A.~Cassan \inst{1}, A.~Cole\inst{5},  J.~Lunine\inst{6,7}, J.~B.~Marquette \inst{1}, S.~Dong\inst{8,9}, A.~Gould\inst{10}, T.~Sumi\inst{11}, V.~Batista\inst{1,10}, P.~Fouqu\'e\inst{12}, S.~Brillant\inst{2}, S.~Dieters\inst{1,5}, C.~Coutures\inst{1}, J.~Greenhill\inst{5}, I.~Bond\inst{13}, T.~Nagayama\inst{14},  A.~Udalski\inst{15}, E.~Pompei\inst{2}, D.E.A.~N\"urnberger\inst{2}, J.B.~Le Bouquin\inst{2, 16}}
          
 \institute{Institut d'Astrophysique de Paris, UMR~7095 CNRS -- Universit\'{e}
    Pierre \& Marie Curie, 98 bis blv Arago, 75014 Paris, France
  \and {European Southern Observatory, Casilla 19001, Vitacura 19, Santiago, Chile}
\and {University College of London, Deparment of Physics and Astronomy, Gower Street, London, WC1E 6BT, UK}
 \and{University of Notre Dame, Department of Physics, 225 Nieuwland Science Hall Notre Dame, USA} 
\and{University of Tasmania, School of Mathematics and Physics, Private Bag 37, GPO Hobart, Tas 7001, Australia}
\and{Dipartimento di Fisica, Universit\`a degli Studi di Roma "Tor Vergata," Rome, Italy}
\and{Department of Astronomy, 610 Space Sciences Building, Cornell University, Ithaca, NY 14853}
  \and{Sagan Fellow}
 \and{Institute for Advanced Study, Einstein Drive, Princeton, NJ 08540, USA}
 \and{Department of Astronomy, Ohio State University, 140 W. 18th Ave., Columbus, OH 43210, USA}
 \and{Department of Earth and Space Science, Osaka University, Osaka 560-0043, Japan}
 \and{Observatoire Midi-Pyr\'en\'ees, UMR 5572, 14, avenue Edouard Belin, 31400 Toulouse, France}        
\and{Institute of Information and Mathematical Sciences, Massey University, Private Bag 102-904, North Shore Mail Centre, Auckland, New Zealand}
\and{Department of Physics and Astrophysics, Faculty of Science, Nagoya University, Nagoya 464-8602, Japan}
\and{Warsaw University Observatory. Al. Ujazdowskie 4, 00-478 Warszawa, Poland}
\and{Laboratoire d'Astrophysique de Grenoble, UMR 5571 Universit\'{e} Joseph Fourier/CNRS, BP 53, F-38051 Grenoble cedex 9, France}
}
\date{}
  \abstract
 	 % {}	 leave it empty if necessary  
{Microlensing is a unique  {method} to probe low mass exoplanets beyond the snow line.
However, the scientific potential of the new microlensing planet discovery is
often  unfulfilled due to lack of knowledge of the properties of the lens and source stars.  The 
discovery light curve
of the super Earth \planet~  suffers from significant degeneracies that limit what can be inferred
about its physical properties.
}
{ High resolution adaptive optics images allow us to solve this problem by resolving the microlensing target
from all unrelated background stars, yielding the unique
determination of magnified source and lens
fluxes. This estimation permits the solution of our microlens model for the mass of
the planet and its host and their physical projected separation. }
{ We observed the microlensing event \event~ at high angular resolution in JHKs 
with the NACO adaptive optics system on the VLT while the object was still amplified by a factor 1.23 and 
then at baseline 18 months later.
We analyzed and calibrated the NACO photometry in the standard 2MASS system in order to 
accurately constrain the source and the lens star fluxes.}
{We detect light from the host star of \planet~, which significantly
reduces the uncertainties in its characteristics as compared to earlier
analyses. We find that \host ~is most likely a very low mass late type
M-dwarf ($0.084 ^{+0.015}_{-0.012}~ \msun$) at a distance of
$660^{+100}_{-70}~\rm{pc}$ orbited by a $3.2^{+5.2}_{-1.8}~\mTer$
super-Earth at $0.66^{+0.51}_{-0.22}~\rm{AU}$. We then discuss the properties of
this cold planetary system. }
   {}

\keywords{-techniques: microlensing, image processing
-exoplanets: individual: \planet, -stars: low-mass stars, late-type}
%  \authorrunning{D.~Kubas, J. ~P. Beaulieu, D.~P.~Bennett, A.~Cassan et al}
\authorrunning{D.~Kubas, J. ~P. Beaulieu et al.} 
\titlerunning{NACO constraints \planet}
   \maketitle
%
%________________________________________________________________

\section{Introduction}

Gravitational microlensing provides a unique window on extrasolar
planetary systems with sensitivity to cool planets, particularly
those of low mass
\citep{1996ApJ...472..660B,2006Natur.439..437B, 2006ApJ...644L..37G,
2008ApJ...684..663B,2008A&A...483..317K, 2010ApJ...710.1641S} that are
currently well beyond the reach of other methods.  Microlensing is
also sensitive to planets orbiting very faint stars and hence spectral
types not routinely examined with other techniques. In general it is a powerful tool to study the Galactic planetary population as a whole \citep{2012Natur.481..167C}.

Microlensing occurs when a foreground (lens) star passes close to the
line of sight towards a background (source) star. The gravity of the
foreground star acts as a magnifying lens, increasing the apparent
brightness of the background star as it gets close to the line of
sight.  A planetary companion to the lens star will induce a
perturbation to the microlensing light curve with a duration that
scales with the square root of the planet mass, lasting typically a
few hours for an Earth to a few days for a Jupiter
\citep{1992ApJ...396..104G,1991ApJ...374L..37M,1964PhRv..133..835L}. Hence
planets are now routinely discovered by dense photometric sampling of
ongoing microlensing events.  The inverse problem, finding the
properties of the lensing system from an observed light curve, is a
complex non-linear one within a wide parameter space.  The planet/star
mass ratio and projected star-planet separation can usually be
measured with high precision.  However in the absence of higher order
effects such as parallax motion and/or extended source effects, in
general there are no direct constraints on the physical masses and
orbits of the planetary system. In the least information case, model
distributions for the spatial mass density of the Milky Way, the
velocity distribution of potential lens and source stars, and the mass
function of the lens stars are used in a Bayesian analysis to
derive probability distributions for the masses of the planet and the
lens star and their distance, as well as the orbital radius and period of
the planet.

With complementary high angular resolution observations, currently
done either by HST or with adaptive optics, it is possible to get
additional strong constraints on the system parameters and determine
masses to about 10\%. This can be done by directly measuring the
light coming from the lens and measuring the lens and source relative
proper motion
\citep{2006ApJ...647L.171B,2007ApJ...660..781B,2010ApJ...713..837B,
2008Sci...319..927G, 2009ApJ...695..970D,2010ApJ...711..731J}

An extrasolar planet with a best-fit mass ratio of $q \sim 2 \times
10^{-4}$ was discovered in the microlensing event MOA 2007-BLG-192
\citep{2008ApJ...684..663B} {\change found by the MOA collaboration
toward the Galactic bulge, (J2000: RA, Dec ) = (18:08:03.8,$-$27:09:00).}
The best fit microlensing model shows
both microlensing parallax and finite source effects.  Combining
these, we obtained the lens masses of $M_l =
0.06^{+0.028}_{-0.021}˜M_\odot$ for the primary and
$3.3^{+4.9}_{-1.6}˜M_\oplus$ for the planet.  The incomplete light
curve coverage of the planetary anomaly led to a significant
degeneracy in the lens models{\bf ,} and the lack of strong constraints on
the source size to a poorly determined Einstein radius.  Together this
resulted in rather large uncertainties in the physical parameter
estimates of the system.

Additional constraints are required to exclude competing microlens
solutions and to refine our knowledge of the physical parameters of
the system.  It is possible to constrain masses and parameters of the
system thanks to high angular resolution imaging. Most microlensing
events only provide a single parameter, the Einstein ring crossing
time $t_E${\bf ,} that depends on the mass of the lens system $M_L$, its
distance $D_L$, the source distance $D_S$ and their relative
 velocity. However, when the relative lens-source proper motion
$\mu_{\rm rel}$ can be determined this yields the angular Einstein
ring radius $\theta_E= \mu_{\rm rel}t_E$.  Moreover $\theta_E$ is
linked to the lens system mass by
\begin{equation}
M_L = {c^2\over 4G} \theta_E^2 {D_S D_L\over D_S - D_L}\ ,
\label{eq-mdl1}
\end{equation}
Therefore, since the distance of the source $D_S$ is known from its
magnitude and colors, Eq. (\ref{eq-mdl1}) is a mass-distance
relation for the lens star. Another constraint is needed to obtain a
complete solution to the microlensing event.  This can be achieved by
directly detecting light from the planetary host star (the lens).
Combining this measurement with Eq. (\ref{eq-mdl1}) and a mass
luminosity relation will yield the mass of the lens.  This has been
done already for several microlensing events where the system is
composed of a star and a gaseous planet
\citep{2004ApJ...606L.155B,2006ApJ...647L.171B,2005ApJ...628L.109U,
2009ApJ...695..970D, 2008Sci...319..927G,2010ApJ...711..731J}.\\ We
observed \event~ in JHK using adaptive optics on the VLT while it was
still amplified by a factor of 1.23 and again when the microlensing
was over. Here, we  combine the NACO JHK flux measurements at these 2
epochs with the color estimate of the source star
\citep{2010ApJ...710.1800G} and the microlensing model
\citep{2008ApJ...684..663B} to disentangle the flux coming from the
source and from the lens star to refine estimates of the parameters of
the system.

%__________________________________________________________________

\section{The data set}

We obtained JHKs measurements using the NACO AO system 
{\change \citep{2003SPIE.4841..944L,2003SPIE.4839..140R} } mounted on
Yepun during the night 6/7 Sept. 2007, while the source star was still
magnified by a factor of 1.23.  AO corrections were performed on a
natural guide star \footnote{The LGSF, which in theory should have
yielded better performance, was not available at that time. However
the pro and contra of LGS vs NGS for us have to be evaluated on a case
by case base, since in the crowded field of microlensing targets 
one often finds suitable NGS references which may give even better
corrections than the LGSF according to the ETC observation preparation
software.}  and observations with the S27 objective ($27'' \times
27''$ FOV, pixelscale=$0.02715''$) were conducted in jitter mode with
multiple exposures at random offsets within $10''$ of the target. In
the absence of suitable "empty" sky patch close to the target, this
strategy was chosen to ensure an accurate estimation of the sky
background and to filter out bad pixels. The second epoch(s) were
obtained with the same observing strategy more than 22 months later
with the event being at baseline, i.e. when the source was not
magnified anymore.  An overview of the NACO data set is given in
Table~\ref{NACOdatable}.\\ To perform absolute calibration of the NACO
images we obtained $90 \times 10$ s dithered images in JHKs 
of the \event\ field with the
Sirius simultaneous 3-band camera \citep{2003SPIE.4841..459N} on
the Japanese/South African IRSF 1.4 m telescope \citep{2000MNSSA..59..110G} 
at SAAO (non AO, $\sim
8'\times 8'$ FOV, pixelscale=$0.45''$) on 29th of
Aug. 2008, i.e. at a time when the event was at baseline.
\subsection{Reduction}
 
Following a "lucky imaging" approach we visually inspect each of the
NACO raw images and remove the ones for which the AO correction was
obviously poor. The remaining raw frames are then dark-subtracted with
darks of exposure times matching the science frames, flatfielded with
skyflats, median co-added and sky-subtracted using recipes from the
Jitter/Eclipse infrared data reduction package by
\cite{1997Msngr..87...19D,1999ASPC..172..333D}. To avoid border
effects, we keep only the intersection of the different dither
positions of the co-added frames for our photometric analysis.\\ The
IRSF data, which was taken to gauge our NACO data, has been
dark-subtracted, flat-fielded and sky subtracted using the
on-the-mountain pipeline package for the SIRIUS camera
\citep{2003SPIE.4841..459N}.

%\begin{landscape}
\begin{table}
\caption{Log of JHKs NACO data. According to the Paranal night logs
the Epoch 1 night was classified as photometric, whereas the Epoch 2
observations were taken in clear sky condition. We give the exposure
time, modified Julian Date, airmass and measured full width at half
max on the coadded frames.}\label{NACOdatable}
\centering
\begin{tabular}{llllcl} 
\hline      
Band & n $\times$~Exp~[s] & MJD &  Airmass&  FWHM [$''$]  &\\%  strehl [$\%$]\\
\hline
  \multicolumn{6}{c}{\it Epoch 1}\\ % To combine 6 columns into a single one
\hline
  J & $6 \times 60$ & 54350.00781250 &1.005&  0.14&\\  %1.6 \\ 
  H & $20 \times 25$ & 54350.02734375 &1.023& 0.19&\\ % 2.2\\
  Ks & $10 \times 25$ &  54349.98828125&1.002&0.09 &\\ %7.2\\
  
\hline
  \multicolumn{6}{c}{\it Epoch 2} \\
\hline
  J &  $23 \times 60$& 55036.08593750 &  1.015& 0.34  & \\
  H &$22 \times 30$ & 55036.06640625& 1.034      & 0.29 & \\
  Ks & $24 \times 30$  &55015.10156250 &  1.088& 0.10  & \\
\hline
\end{tabular}
\end{table}
%\end{landscape}

 %
\begin{figure}
   \centering
     \includegraphics[width=9cm]{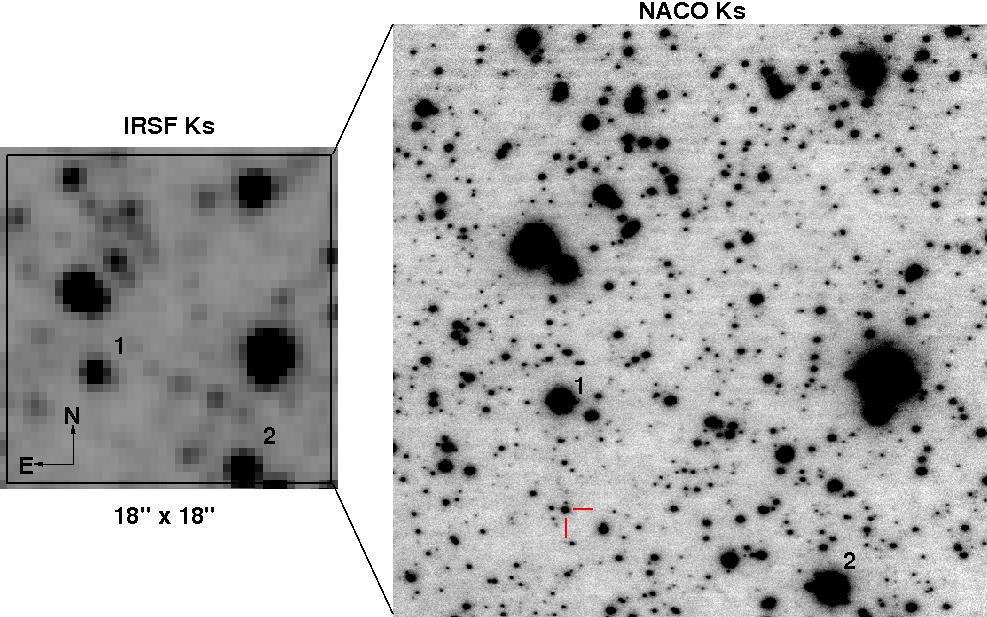}

\caption{{\bf Left:} Extract of IRSF Ks band image of \event~ used to
calibrate the NACO photometry of the $18''\times 18''$ large
intersection fov of the coadded NACO frames in Ks band ({\bf right}).
\event~is marked with the half cross hair. The stars annotated with
"1" and "2" serve as PSF-reference and photometric zeropoint
calibrators. Furthermore these two stars are common to 
all bands and epochs. The bright stars north of the two references are either
too crowded, in the non-linear regime or too far away from the target. }
              \label{FigNACO_IRSF_CALIB}
\end{figure} 

%%%%%%%%%%%%%%%%%%%%%%%%%%

  \begin{figure*}
   \centering
     {\includegraphics[ width=9cm]{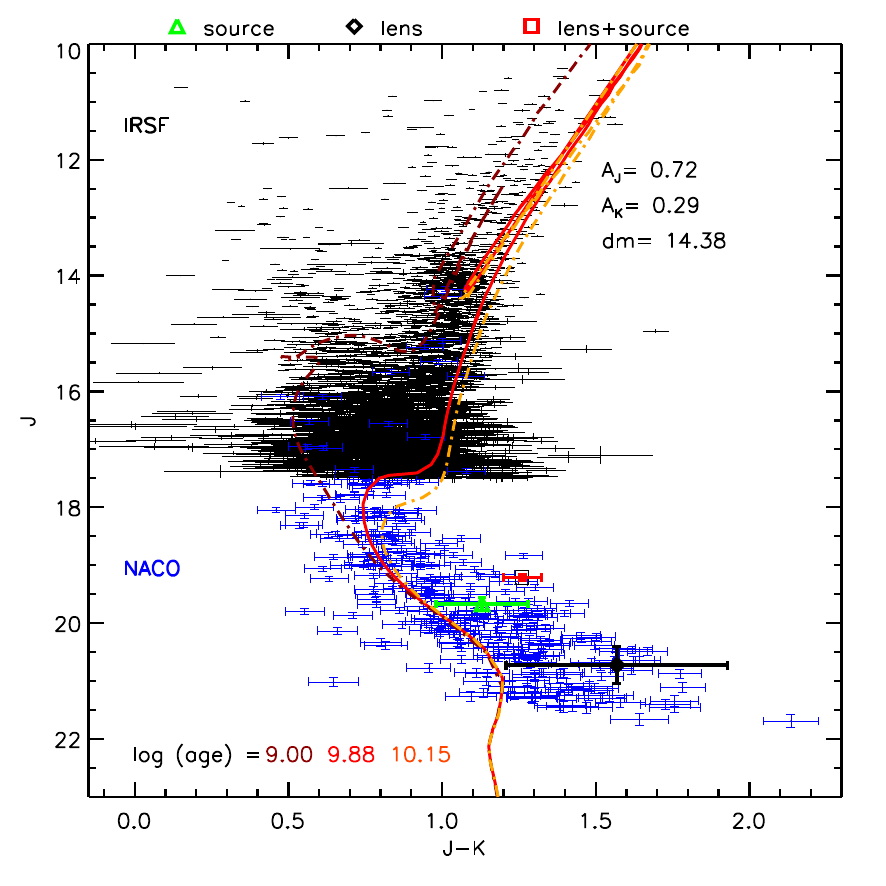}}
    { \includegraphics[ width=9cm]{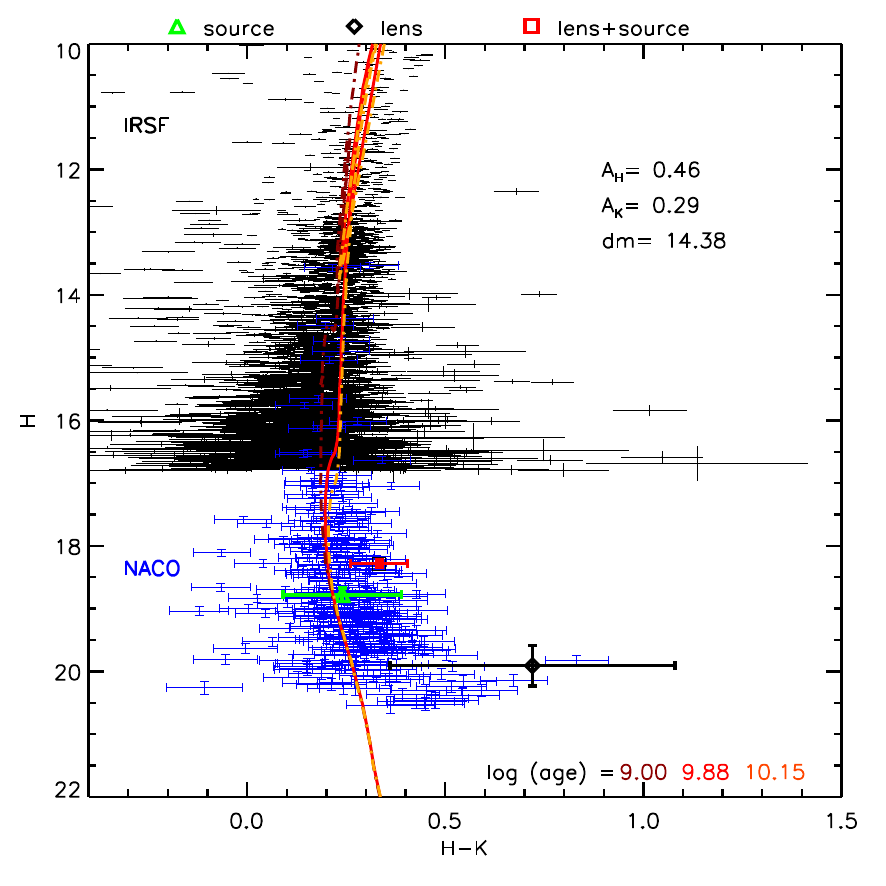} }

\caption{{\bf Left}: The (J-Ks, J) CMD in the 2MASS system of the
\event ~ field combining the data from the IRSF (within $3'$ of
target, black points) and NACO (within $18''$, blue points). In red
the photometry of the measured lens+source flux at magnification
A=1.23 is displayed together with the inferred decomposed fluxes of
the source (green) and the lens (planetary host star,
black). Overplotted are Marigo et al (2008) solar metallicity
isochrones of ages $\log ({\rm Gyrs})= 9.00, 9.88,10.15$ at distance
modulus of dm=14.38 and estimated extinction of $A_J=0.72,
A_{Ks}=0.29$. {\bf Right}: Same as above but for (H-Ks, H).  }
              \label{FigCMDs_1}%
    \end{figure*}
  
    \begin{figure}
   \centering
    { \includegraphics[ width=9cm]{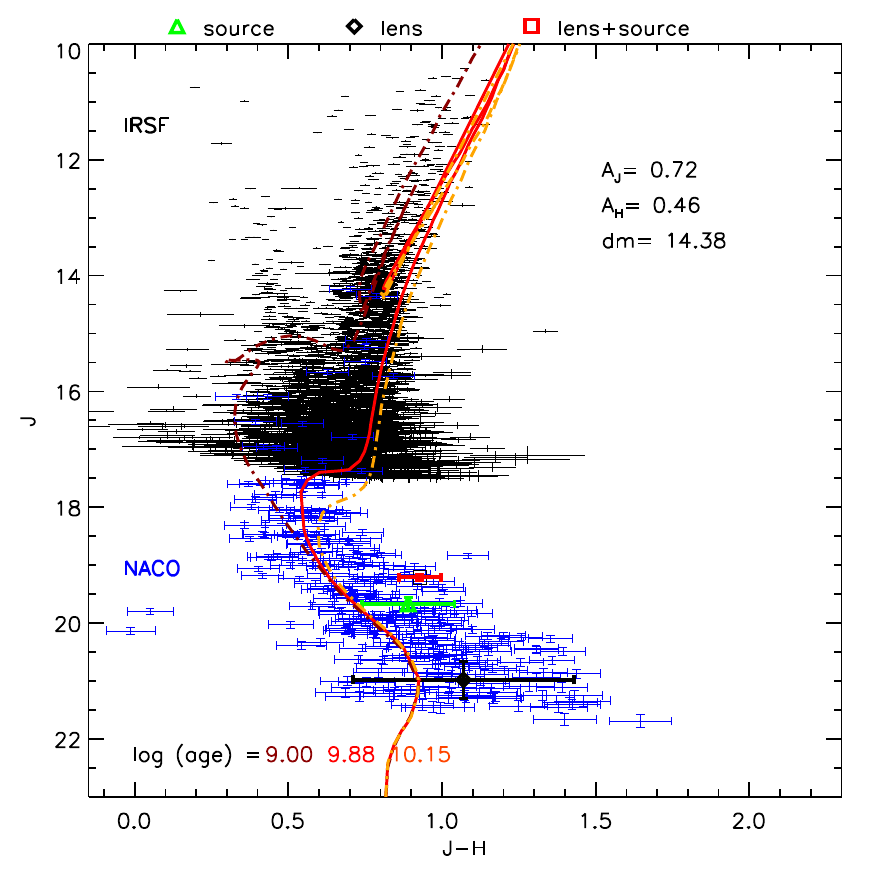}}
     \caption{Same as Fig.~\ref{FigCMDs_1} but for   (J-H,J).
   }
              \label{FigCMDs_2}
    \end{figure}

\section{Photometric Analysis}
As in our previous analysis of planetary microlensing event
\planetjulia ~\citep{2010ApJ...711..731J}, we extract the photometry
of NACO images using Starfinder \citep{2000A&AS..147..335D}. This tool
is tailor suited to perform photometry of AO images of crowded
fields. It creates a numerical PSF template from chosen stars within
the frame, which is then used for PSF-fitting of all stars in the
field.  To build our PSF reference we chose the star marked as ``1'' in
Fig.~\ref{FigNACO_IRSF_CALIB} based on the following criteria. It is
close to the target (within less than $4''$), sufficiently bright but
well within the linearity regime of the detector and common to all
final reduced JHKs images of both epochs. Fig.~\ref{psfsub} shows the
JHKs images centered on the target for the first epoch of NACO 
and the PSF subtracted residuals.
The IRSF photometry catalog was created with DoPhot \citep{1993PASP..105.1342S}

%%%%%%%%%%%%%%%%%%%%%%%%%%

\subsection{Building a calibration ladder}\label{ladder} 

In order to build the calibration ladder, we  use three data sets:
the 2MASS catalogue, photometry obtained at the IRSF telescope and then NACO data.
It is necessary to use the intermediate step of IRSF observations because we have too
few stars in common between 2MASS and NACO. Such ladder has been used already 
and described in the appendix of \citep{2010ApJ...711..731J}. 
The three JHKs color systems are very close. For example, the color term for J-Ks colors 
between 2MASS and IRSF is only 0.01.  Accurate calibration between 2MASS
and IRSF has been given in \cite{2007PASJ...59..615K}.  \cite{2010ApJ...711..731J} 
did not detect color terms between NACO and IRSF filters. Therefore, our calibration
ladder must only determine the zeropoint offsets.

We first perform the astrometry of the IRSF images with respect to the
online 2MASS catalog using GAIA/Skycat and WCSTools.  Then, using only
stars marked as AAA (highest 2MASS quality flag) in the JHKs bands we
crossmatch the common stars to compute the photometric transformation
between the two catalogs by sigma clipping, demanding an astrometric
accuracy of the match of better than $0.6''$. To
minimize the effect of source confusion and blending contamination we
cut off at magnitude 13 for the 2MASS reference stars and sum up the
flux of close neighbors for the IRSF sources to account for the much
coarser pixel scale of the 2MASS catalog.

The PSF reference star is contained in the IRSF catalog, as well as
star ``2'' (Fig.~\ref{FigNACO_IRSF_CALIB} ).  We examine their long term
photometric stability in the OGLE database and find that over more
than seven years both stars are stable (in the optical $I$-band) at
levels of $\lesssim 1\%$, which makes them well suited as zeropoint
calibrators of our NACO field. While we adopt star ``1'' as 
the primary
photometric calibrator since star ``2'' is more crowded, we determine
zeropoints from both stars as a consistency check. To account for the
different plate scales between NACO and IRSF we sum up the flux of all
the NACO sources which are contained within the IRSF PSF.  We note
that observing conditions (sky transparency and atmospheric coherence
times) for the second epoch data set were inferior to the epoch 1
measurements and the uncertainties in the absolute zeropoints of epoch
2 are therefore larger.  Since we are mainly interested however in the
relative photometry of the two epochs we can align the epoch 2
photometry with respect to more accurately calibrated epoch 1.  Table
\ref{ZPtable} summarizes this way of determining the transformations
to calibrate the NACO data with respect to the 2MASS system and Table
\ref{TARGETtable} shows our derived photometry for \event.

\begin{table}
\caption{JHKs NACO photometry for \event, i.e. lens+source (no
dereddening applied).  The absolute photometry error budget is
composed by adding in quadrature the errors on the zeropoint, the
formal error reported by Starfinder and the background error as
estimated from the scatter between epoch 1 and epoch 2 comparison
stars. For epoch $1$, $J$, and $H$ bands, we adopt the background error
estimate as derived from the $K$ band, since the poor epoch 2 quality in
J and H would overestimate the epoch 1 errors.  }
\label{TARGETtable}
\centering
\begin{tabular}{ccccl} 
\hline       
Band & $J$ & $H$ & $K_\mathrm{s}$   \\
\hline
  \multicolumn{4}{c}{ {\it {\bf NACO Epoch 1}} }\\ % To combine 6 columns into a single one
\hline
 \hline
    \multicolumn{4}{l}{\it ~calibrated against IRSF  $~~~~$   } \\
\hline
   & $19.209\pm 0.043 $  &  $18.281\pm 0.042 $  &$17.948\pm 0.035$  \\ 
  \hline
  \multicolumn{4}{c}{\it {\bf NACO Epoch 2}} \\
  \hline
  \multicolumn{4}{l}{\it ~calibrated against IRSF $~~~~$} \\
\hline
 & $19.324\pm 0.073 $  &  $18.548\pm 0.112 $  &$17.989\pm 0.038$  \\ 
  \hline
  $\Delta {~\rm Epochs} $ & $0.115\pm 0.085 $  &  $0.267\pm 0.120 $  &$ 0.041 \pm 0.052$  \\ 
  \hline \hline
   \multicolumn{4}{l}{\it ~aligned with respect to Epoch 1 } \\
   \hline
& $19.283\pm 0.071 $  &  $18.498\pm 0.087 $  &$18.011\pm 0.042$  \\
\hline
 $\Delta {~\rm Epochs} $ & $0.074\pm 0.083 $  &  $0.217\pm 0.097 $  &$ 0.063\pm 0.055$  \\ 
\hline
\end{tabular}
\end{table}

\begin{figure*}
  % \centering
    {\includegraphics[width=6.cm]{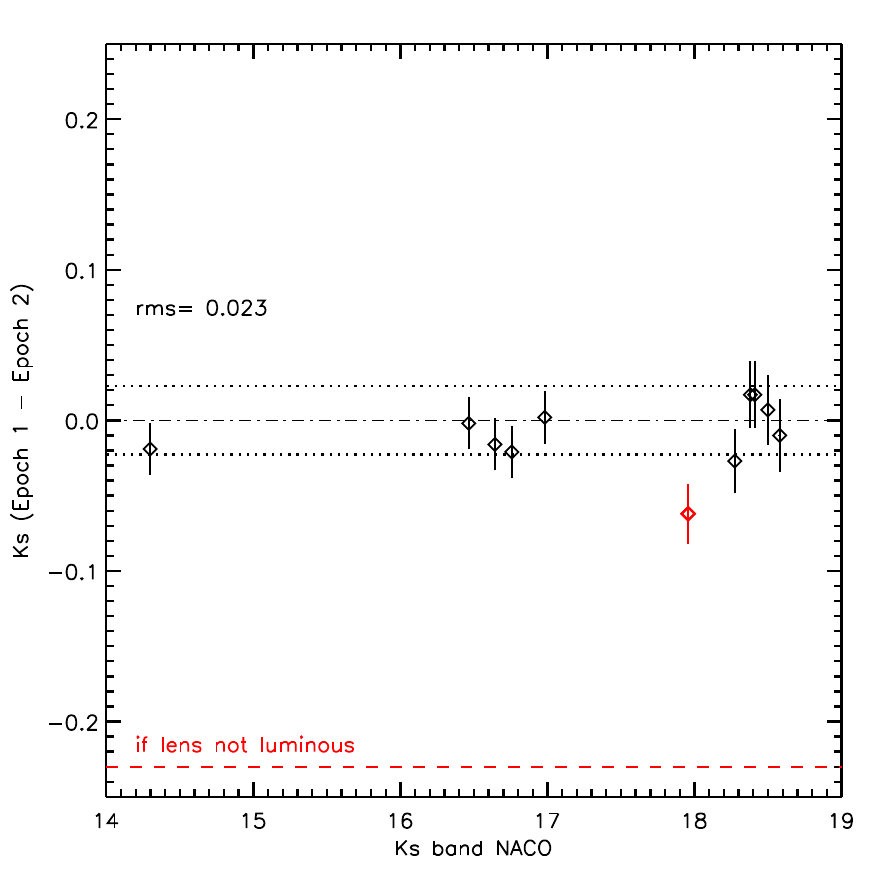}}
   {\includegraphics[width=6cm]{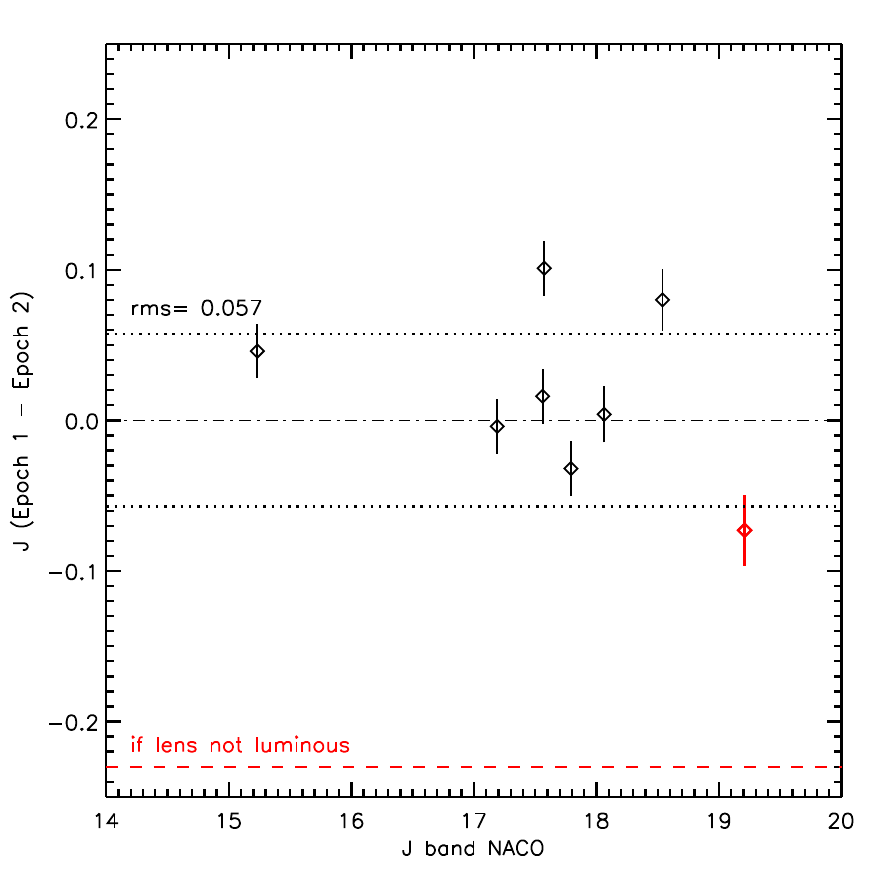}}
  {\includegraphics[width=6cm]{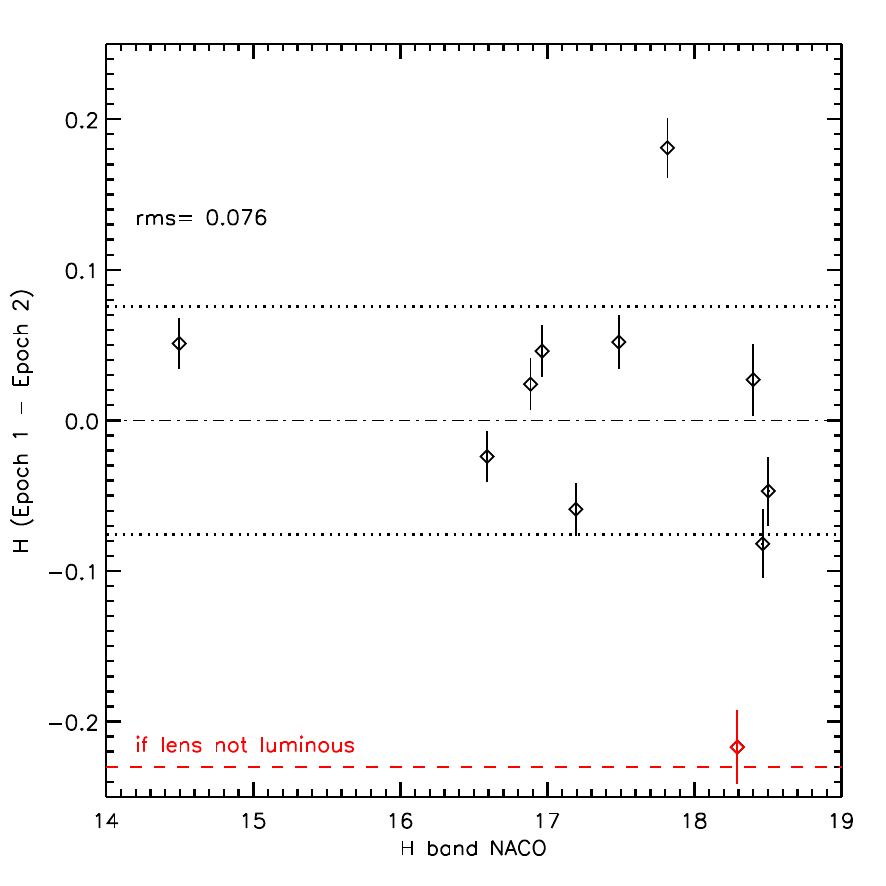} }
\caption{ Relative photometry of the NACO epochs for each band
calibrated / aligned to Epoch 1 based on stars within $4''$ of
\event. {\change The black points are constant stars, so their
scatter gives an estimate of the error.  
The red dashed line marks the expected magnitude difference of
the target assuming no light from the lens is detected. 
The red diamonds show the flux changes of the target, which are} 
inconsistent with such
a scenario at a $3-\sigma$ level for J and even with higher
significance for our best data set in $K$ band. The poor quality of the
epoch 2 $H$ band data as evident by the large scatter does not provide
any useful constraint.  Table \ref{TARGETtable} gives the derived
values for all bands and different choices of the field of comparison
stars.  } \label{Fig:NACOEp1Ep2}
    \end{figure*}

\section{Results}\label{sec:results}

In Fig.~\ref{FigCMDs_1} and ~\ref{FigCMDs_2} we present the
color-magnitude diagrams for the combined IRSF and NACO (epoch 1)
data. To estimate the interstellar extinction, we first determine the
position of the red clump center by taking the median of the
distributions in color and magnitude inside a window centered on a
first guess estimated position.  Then we fit the tip of the Red
Giant Branch as given by the isochrones of \cite{2008A&A...482..883M}
adopting the distance modulus $\rm{dm}=14.38\pm 0.07$ as found for the
\event~field by \cite{2008ApJ...684..663B}.  With a best fit age of
$\log({\rm Gr})=9.88 $ we find for the extinction coefficients:
$A_J=0.72 \pm 0.10$, $A_H=0.46\pm 0.10$, $A_K=0.29 \pm 0.10$, which
for this line of sight is consistent with extinction maps from
\cite{1998ApJ...500..525S} and \cite{2006A&A...453..635M}

\subsection{The case for a luminous lens {\change I: NACO-only}}
  The standard general microlens light curve model is given as
 \begin{equation}
 F(t) = F_{\rm S} A(t) + F_{\rm B},
 \label{lc}
 \end{equation} 
where $F$ is the measured flux at the telescope, $F_{\rm S}$ is the
intrinsic unmagnified source flux, $A(t)$ the time dependent
magnification given by the lens model and the blend flux $F_{\rm
B}=F_{ \rm L}+F_{\rm Background }$, which contains the lens flux $F_{
\rm L}$ and {$F_{\rm Background }$} the flux of any unrelated 
field stars within the aperture{\bf, F$_{\rm Background}$}.  
While the source flux $F_{\rm S}$ can be
determined with high precision from the light curve modeling of the
non-AO data, given a large magnification gradient, the background term
normally dominates over the lens term in seeing-limited photometry
of typically crowded Galactic Bulge fields of microlensing.  Hence the
benefits of high spatial resolution imaging 
are obvious. Reducing or eliminating the contribution of
contaminating background sources provides a better estimate of the
lens flux and so finally of the physical characteristics of the lens
system. In \cite{2010ApJ...711..731J} the lens flux could be estimated
by comparing a single NACO AO epoch with an excellent seeing-limited
light curve in the same passband from which the source flux had been
previously determined with good accuracy.  For \planet~we have no such
light curve in the NACO passbands but a well determined measurement of
the source flux in the $I$ band (Cousins system), which we can transform
into the expected source flux for JHKs bands.  Note that while in
theory our two point NACO "light curve" can be used to solve
Eq.~(\ref{lc}) for the lens and source fluxes directly, the resulting
uncertainties are very large \citep{2009ApJ...695..970D} due to very
small magnification "lever arm" for our event and so following the
path of \cite{2010ApJ...711..731J} is much more accurate.
 
First, however, the two epochs can be used as follows, without the knowledge of
the source flux{\bf, } to check whether there is an indication 
that light from the lens is detected.
The expected magnification gradient
between the two NACO epochs based on the best-fit model of
\cite{2008ApJ...684..663B} is $\Delta m = 0.230 \pm 0.015$~mag. Note
that this gradient is basically the same for all competing planetary
models, since the first epoch was taken close to the baseline of the
event, where the single lens approximation describes the data very
well.  If the lens is dark and no unrelated source is contaminating
our photometry (see Sec.~\ref{alt}) we then would expect to measure
this difference in the relative photometry of the two epochs in each
band. Since the quality of epoch 1 is superior we choose epoch 1 as
reference to which we align epoch 2.  We compare the photometry
between the two epochs for each band using 3 different alignment
procedures. First we compare the derived absolute photometry (with
respect to 2MASS using the calibration ladder described in
Sec.~\ref{ladder}).
 
Then we align epoch 2 with respect to (calibrated) epoch 1 using all
common stars within $4''$ (to minimize effect of PSF variations) of
the target.  The resulting magnitude differences for the target and
the absolute photometry values are summarized in
Table~\ref{TARGETtable}.  The difference between the epochs is shown
in Fig.~\ref{Fig:NACOEp1Ep2}. Regardless of the 
alignment method used, for all bands except H (the set with the poorest epoch 2 data
quality), the measured difference is less than in the case of a dark
lens, albeit with different levels of significance.  For $K$ band, the
best data set, a dark lens is inconsistent with the measurement at $2
\sigma$ for the absolute alignment and at $3 \sigma$ for the relative
alignment.  {\change The results for $J$ band are also inconsistent with 
a dark lens, but in this case at slightly less than $2\,\sigma$.}
This motivates a more careful examination of the evidence for
a luminous lens.

{\change
%\vspace{6cm}
\hspace{-9cm}
\begin{figure}
\centering { \includegraphics[ width=9.9cm]{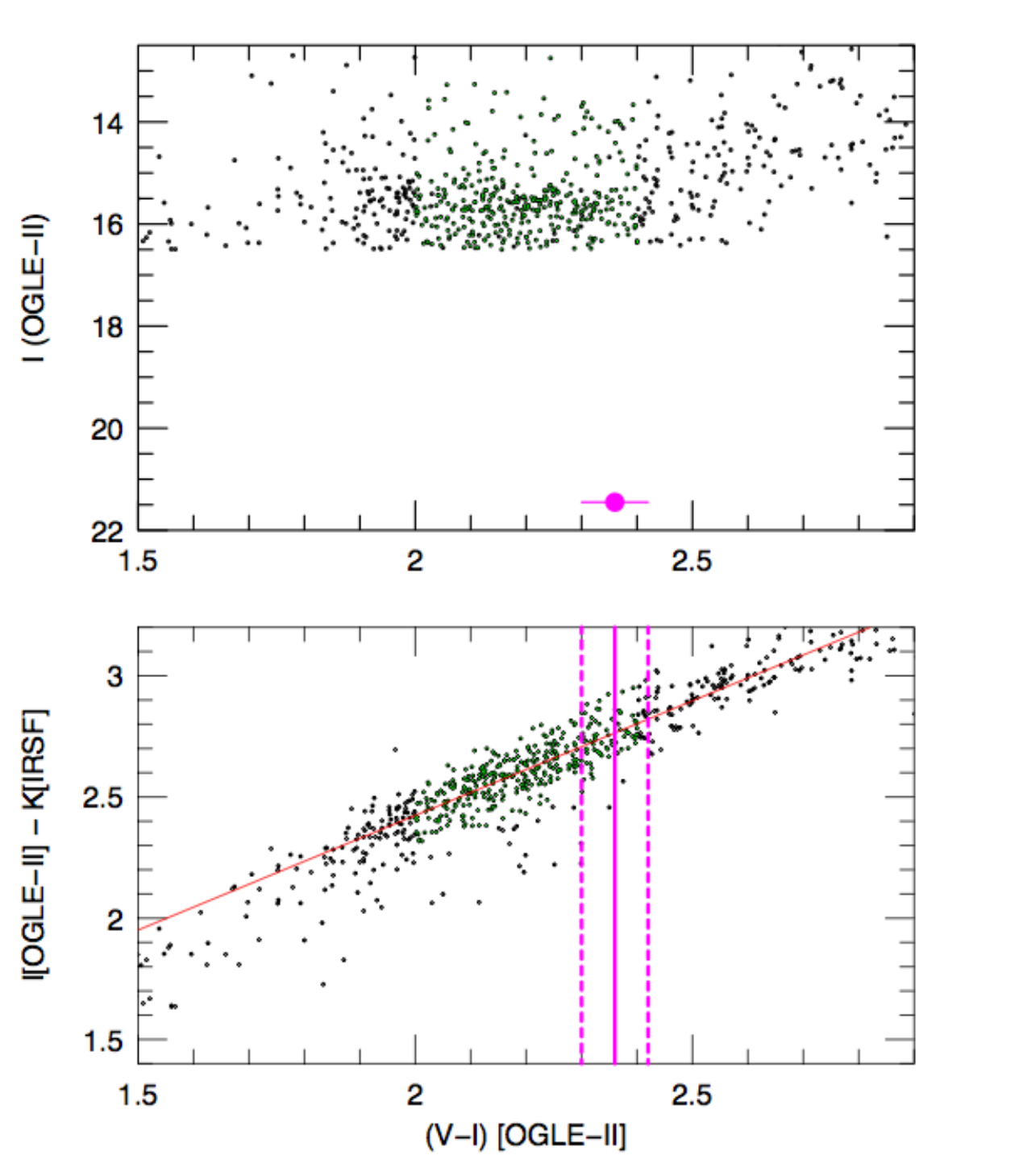}}
\caption{{\change Empirical $VIK$ instrumental color-color relation. 
Lower panel: Open black symbols show all astrometric matches
(that survived a strict crowding criterion)
between OGLE-II $V/I$ data and instrumental IRSF $K$ data.
The green filled points show those used in the fit (red line),
which are restricted to $2.0<(V-I)<2.4$ and exclude $3\,\sigma$
outliers. Vertical magenta lines show MOA-2007-BLG-192S $(V-I)$
color measurement by \citet{2010ApJ...710.1800G}.  Upper panel
shows the $VI$ CMD positions of stars used to determine the $VIK$
relation.  They are all giants in or near the clump, whereas the
source (magenta point) is a dwarf.  Nevertheless,
\citet{1988PASP..100.1134B} show that the $VIK$ relations are
essentially identical for giants and dwarfs in this color range (see text).}}
\label{colcol}
\end{figure}

\subsection{The case for a luminous lens II: NACO+IRSF+Optical}

A more powerful test for the presence of ``excess light'' (in addition
to that of the source) is possible by combining NACO, IRSF, and optical data.
To maximize sensitivity, we will work
entirely with {\it uncalibrated} data.  This will eliminate  any errors
associated with calibration relations, extinction estimates, and color-color relations.
The remaining errors, which are either measurement errors or 
intrinsic scatter, are both small and easy to measure.

We begin by constructing a color-color diagram that
combines optical $V$ and $I$ data from OGLE-II with $K$ data from
IRSF  (see Fig.~\ref{colcol}).  The OGLE-II data are used because this
is the system in which \citet{2010ApJ...710.1800G} measured the color and
magnitude of MOA-2007-BLG-192S, 
\begin{equation}
(V-I)_s=2.36\pm 0.03; \qquad I_s=21.45\pm 0.05.
\label{eqn:visource}
\end{equation}
The OGLE-II data are in fact calibrated, but that is incidental: the
important thing is that the optical color is measured in this system.
The $K$ magnitudes are constructed directly from IRSF photometry
fluxes $K=22.155 - 2.5\log({\rm flux})$.  The zero-point constant
is chosen for convenience to be similar to the calibration constant,
but this constant does not enter the calculation in any way.  In 
particular, the data remain uncalibrated because there is no
color term.  

The open black circles are all the astrometric matches that meet a
strict crowding criterion.  A color-color relation (red) is derived
by fitting the points in the range $2.0<(V-I)<2.4$ with $3\,\sigma$
rejection (green filled points).  This choice of interval will be
justified below.  The relation is:
\begin{eqnarray}
(I-K) =  Z_1 + Z_2[(V-I)-2.36];\nonumber\\
Z_1 =  2.757 \pm 0.008;
\quad
Z_2 = 0.943\pm 0.039
\label{eqn:vikregress}
\end{eqnarray}
with a scatter of 0.080 mag.

The vertical lines represent the best fit and error bar of the
Gould et al.\ (2010) optical-color measurement.  From Eq.s 
(\ref{eqn:visource}) and (\ref{eqn:vikregress}),
the best estimate of the source magnitude at the first epoch
(when the source was magnified by $A=1.23$) is therefore
\begin{equation}
K_s =  [I_s - 2.5\log(A)] - Z_1 - Z_2[(V-I)_s-2.36] 
=   18.468
\label{eqn:kseval}
\end{equation}

We discuss all the errors in this estimate below.

We now compare this with the baseline flux as measured by NACO
and transformed to {\bf the} IRSF {\bf system:}
%\begin{equation}
\begin{eqnarray}
K_{\rm base,IRSF} &=& \nonumber\\
K_{\rm comp,IRSF} 
   &+& 2.5\log(F_{\rm comp,NACO}/F_{\rm base,NACO}) = 17.948
\label{eqn:kbaseeval}
%\end{equation}
\end{eqnarray}
where $K_{\rm comp,IRSF} = 14.261 \pm 0.016$ is the IRSF magnitude
of the comparison star, and $F_{\rm comp,NACO}=88066\pm 37$
and $F_{\rm base,NACO} = 2952.2 \pm 12.1$ are the NACO
fluxes of the comparison and baseline stars, respectively.

The baseline flux is clearly larger, $K_s - K_{\rm base} = 0.520$,
The question is, how large is the error in this difference?
Eq. (\ref{eqn:kseval}) has 5 identifiable sources of error.  First, the
error in $I_s$ is 0.05, but the error in $[I_s - 2.5\log(A)]$
is actually smaller than this by a factor 0.57 (see Eq.~10 of 
\citealt{2010ApJ...711..731J}).
Second, the error in $Z_1$ is 0.008.  Third the error in the
final term is $Z_2$ times the error in $(V-I)_s$, i.e., 0.056.  

Fourth,
we are using the $VIK$ color-color relation of the field stars observed
by IRSF as a proxy for the $VIK$ color-color relation of the source.
However, the source is a dwarf, while the field stars are all giants.
Now, according to Fig.~1 of \citet{1988PASP..100.1134B}
these $VIK$
relations of giants and dwarfs are virtually identical for $(V-K)_0<3.0$
[$(V-I)_0<1.3$], and diverge only very slowly at redder colors.
The dereddened color of the source is $(V-I)_0=1.24\pm 0.06$, so
the entire $1\,\sigma$ error range lies within the ``same relation''
region.  And again, the relations diverge only very slowly at redder
colors.  We note, however, from Fig.~2 of \citet{1988PASP..100.1134B},
that the divergence is extremely rapid in $(J-K)$.  This is the
principal reason that we conduct this test in $K$ rather than $J$.

Finally, the $VIK$ relation in Fig.~\ref{colcol} 
exhibits a scatter of 0.08 mag.
If this scatter is attributed to measurement errors, then the effect is
very small.  There is some reason to expect that this is the case
because Fig.~1 of \citet{1988PASP..100.1134B}
shows almost zero scatter.
However, the bulge star population may be more diverse than the local
one.  For the moment we assume it is intrinsic, and that this scatter
in the observed giant $VIK$ relation also applies to dwarfs.  Then,
the error in $K_s(A=1.23)$ is 
$[0.028^2 + 0.008^2 + 0.056^2 + 0.08^2]^{1/2} = 0.102$.

There are three errors contributing to $K_{\rm base}$, which are listed
below Eq. (\ref{eqn:kbaseeval}).  Their sum in quadrature is $0.017$.  
As seen in section 3 the JHKs bands from 2MASS, IRSF and NACO 
are very close, with negligible color terms in the transformations
between the different systems. 
Next we note that photometry on AO images is notoriously difficult
due to gradients in the PSF.  This effect is hard to quantify.
However, Fig.~\ref{psfsub} shows that it must be quite small.  The left 
(right) column shows the image before (after) PSF subtraction for the first
epoch, for $J$, $H$, $K$, respectively.  The 
$K$ image in particular, looks extremely clean.  We nevertheless
conservatively estimate a 0.03 photometry error due to PSF gradient.
This yields a total error on $K_{\rm base}$ of 0.034 mag.

Finally, we note that even if the scatter in Fig.~\ref{colcol} were due
to measurement noise, rather than intrinsic scatter (and so should
not have been included in the error in $K_s(A=1.23)$, it would then
contribute to the error in $K_{\rm base}$ through $K_{\rm comp,IRSF}$.
Hence, the impact on the final error would have been identical.

We therefore finally derive our estimate of the excess magnitude
at baseline:
$$
K_s - K_{\rm base} = 0.520 \pm 0.108
$$
which is a $4.8\,\sigma$ detection of additional light. It is either a blend aligned to better 
than 0.1 arcsec with the source star of the microlensing event or light coming from the lens star. 

\begin{figure}
\centering { \includegraphics[ width=9.2cm]{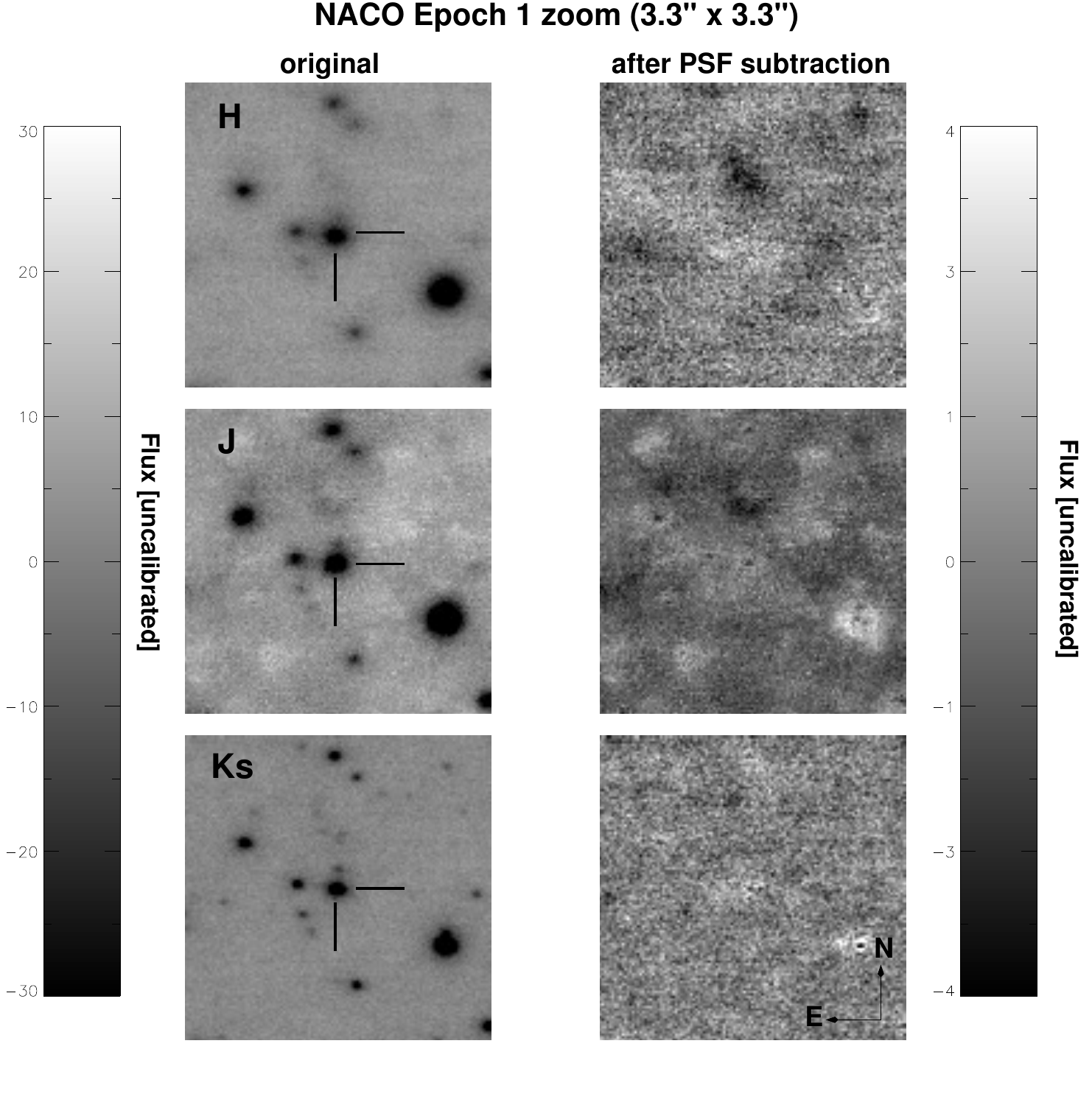}}
\caption{Left: Zoom ($3.3" \times 3.3"$ ) of Epoch 1 NACO images centered on the target in $H$ (top), $J$ (middle), 
and $K$ (bottom).  Right: Images after PSF subtraction. The PSF subtraction does a
good job of removing essentially all flux, particularly in $K$.  
}
\label{psfsub}
\end{figure}

%}

\subsection{Is the blended light from the lens star? }\label{alt}
The mean density of stars of comparable brightness and color $\pm
0.20$~mag to the detected blend is less than 0.2 per ${\rm arcsec}^2$
as derived from our best / sharpest data set, the Ks band of epoch
1. Given the image quality of $0.09''$ FWHM, this conservatively
implies a probability of less than $2 \%$ for the blend being
unrelated to the microlens event.  Another possibility to consider is
that the blend stems from a companion to the source star. Close
companions with periods $\lesssim 100 ~\rm {d}$ can be ruled out by
the xallarap signal limits in the light curve and very wide separation
companions $\gtrsim 700~\rm{AU}$ would be resolved in the Ks NACO
data. This still leaves a large range of allowed separations but
taking into account the color difference the possible fraction of low
mass secondaries should not be larger then $8\%$ according to
\cite{1991A&A...248..485D}.  However only future AO or HST images,
when the source and lens will have moved sufficiently far apart to be
spatially resolved, will be able to securely rule out such a scenario.
   
\subsection{Source star constraints}
{\change To compare the $(V-I)_0=1.24\pm 0.06$ color found by
\citet{2010ApJ...710.1800G}
 to the NIR bands of this study we transform this
$V-I$ color to $J-K$ in the 2MASS system using first the dwarf color table
of \cite{1988PASP..100.1134B} to find $(J-K)_{\rm 0,s}=0.73$ and then with 
the 2MASS-Bessell$~\&~$Brett filter relation\footnote{http://www.astro.caltech.edu/~jmc/2mass/v3/transformations/}
we finally derive $(J-K)_{\rm 0,s}=0.70\pm 0.07$.

From our NACO "light curve" using Eq.~\ref{lc} and our lens model , we
find after dereddening $(J-K)_{\rm 0,s}=0.66\pm 0.51$. While the
uncertainty derived from linear regression is large, this independent
source color determination is very consistent with the colors found by
\cite{2010ApJ...710.1800G} as well as those of \cite {2008ApJ...684..663B} and
strengthens the case for the source being a K4-5 dwarf in the Bulge at
$7.51 \pm 0.25~ \rm{kpc}$.  However, given the better accuracy of the
\cite{2010ApJ...710.1800G} source color, we adopt their value in the
following analysis.}
\subsection{Lens/planetary system  constraints}
From the \event~ light curve the $I$ band source flux is well determined
to $I_{s}=21.44 \pm 0.08$ \citep{2008ApJ...684..663B}.  Using the
source color derived in the previous section and the extinction
coefficients determined from the IRSF data, we can translate this $I$
band estimate into the NACO passbands to derive $J_{s}=19.67 \pm 0.12,
H_{s}=18.78 \pm 0.10,K_{s}= 18.54\pm 0.10$ (2MASS system).
 
Using our best lens model and Eq.~(\ref{lc}) we then derive the
following estimates for the apparent lens flux: $J_{l}=20.98
\pm 0.30, H_{l}=19.91 \pm 0.30, K_{l}= 19.29\pm 0.20$ from Epoch 1 and
$J_{l}=20.59 \pm 0.40, H_{l}=20.10 \pm 0.50, K_{l}= 19.04\pm 0.20$
from Epoch 2.  Taking the weighted average we finally get as best
estimate for the lens flux: $J_{l}=20.73 \pm 0.32, H_{l}=19.94 \pm
0.35, K_{l}=19.16 \pm 0.20 $.

We now can use mass luminosity relations to translate the photometry
estimates of the apparent lens flux into estimates of the planetary
host star mass.  We adopt the relations of \cite{2000A&A...364..217D}
for M-dwarfs (with masses $> 0.10 ~\msun$) and
\cite{2000ApJ...542..464C} for L-dwarfs (masses $<0.10 \msun$), where
the transition between the two relations at $ \sim 0.10~ \msun$ has
been linearly interpolated. The best lens model gives an estimate for
the distance and mass of the lens via the measurement of the parallax
$\pi_E$ using Eq. (\ref{eq-mdl1}). In Fig.~\ref{FigLensmassLight}
the implied apparent lens brightness based on the mass-magnitude
relations and our constraint on the parallax is plotted as a function
of lens mass. All bands agree that the lens mass is in the range $
0.07< M_{\rm L}/ \msun <0.10$ with a best estimate of $M_{\rm L}/
\msun= 0.087 \pm 0.010$, preferring a stellar over a sub-stellar
host. This is consistent with the previous best estimate of $M_{\rm
L}/ \msun= 0.06 \pm 0.04$ \citep {2008ApJ...684..663B}, but which was
not able to distinguish between the different host star
possibilities. This new refined lens mass 
also affects the inferred planetary mass
of \planet.  This is due to a light curve degeneracy between the
planetary mass ratio $q$, and the source star radius crossing time
$t_\ast$. The detection of light from the lens star means that it must
be massive enough to be above the hydrogen burning threshold, which
constrains $t_\ast < 0.05\,$days and rules out the cusp crossing
models (models I-P of Table 1 in \citet{2008ApJ...684..663B}
; the
remaining surviving models consistent with the NACO data are listed
here in the appendix in table \ref{tabsurvive}1). This constraint on
$t_\ast$ pushes the mass ratio, $q$, toward somewhat smaller
values. As a result the range of allowed planetary masses is nearly
unchanged.

The physical parameters of the star-planet system can be estimated by
the same type of Markov Chain Monte Carlo (MCMC) calculations used in
\citet{2008ApJ...684..663B} or \citet{2009ApJ...695..970D}. But we now
add constraints that the lens star must satisfy the JHK mass
luminosity relations of \citet{2000A&A...364..217D}, under the
assumption that 25\% of the dust responsible for the extinction of the
source star is also in the foreground of the lens star plus planet
system. The uncertainty in the lens magnitude is taken to be 0.3 mag
in each passband.  This accounts for the uncertainty in the extinction
as well as the uncertainty in the \citet{2000A&A...364..217D}
mass-luminosity relations, which become large at low masses because of
the metallicity dependence of the minimum stellar mass. The parameter
values resulting from this calculation are listed in
Table~\ref{tab-mcmc}. The planet mass is now $3.2^{+5.2}_{-1.8}\mTer$,
while the host star mass is $0.084{+0.015\atop -0.012} \msun$ and the
two dimensional star-planet separation during the event is $a =
0.66{+0.51\atop -0.22} \,$AU.  The MCMC lens distance estimate is $D_L
= 700{+210\atop -120} \,$pc which agrees with our more direct estimate
of $660^{+100}_{-70}\rm{pc}$. This implies that the lens suffers less than 
half of the total extinction
toward the source, and our derived lens colors are consistent with a late
M spectral type \citep{2010ApJ...710.1627L} of the planetary host.

%
%                                                
%----------------------------------------------------------- LENS MASS CONSTRAINTS
%
% routine Ldwarf_isochrones_X.pro
%
  \begin{figure}
   \centering
  \includegraphics[width=7.1cm]{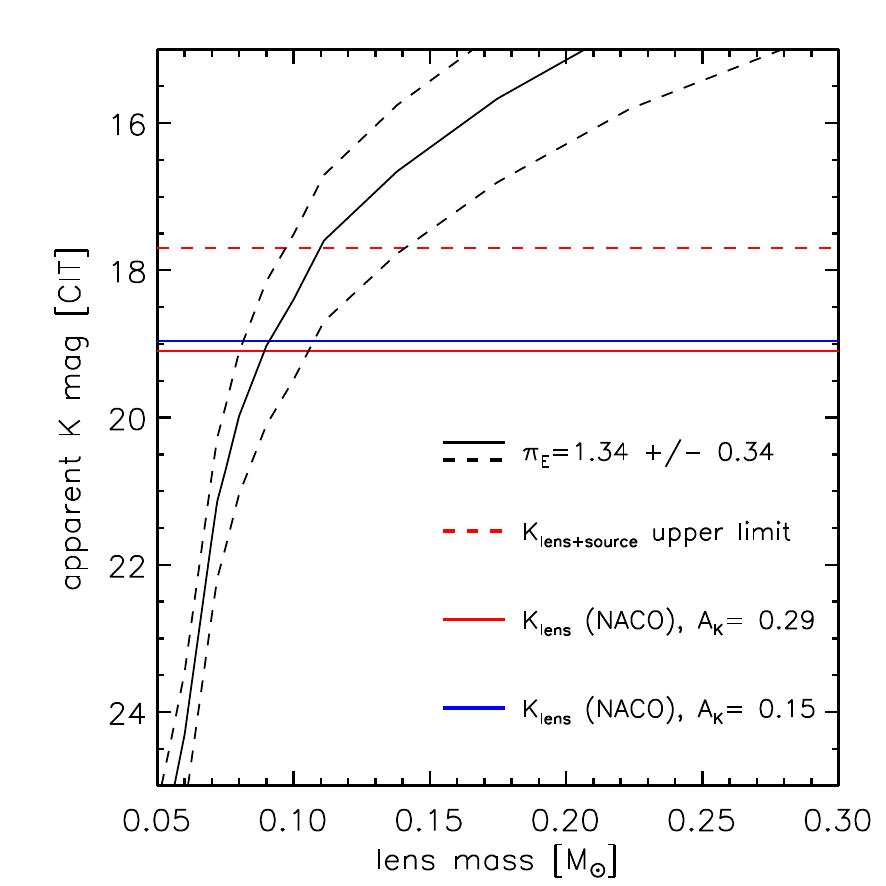}
  \includegraphics[width=7.1cm]{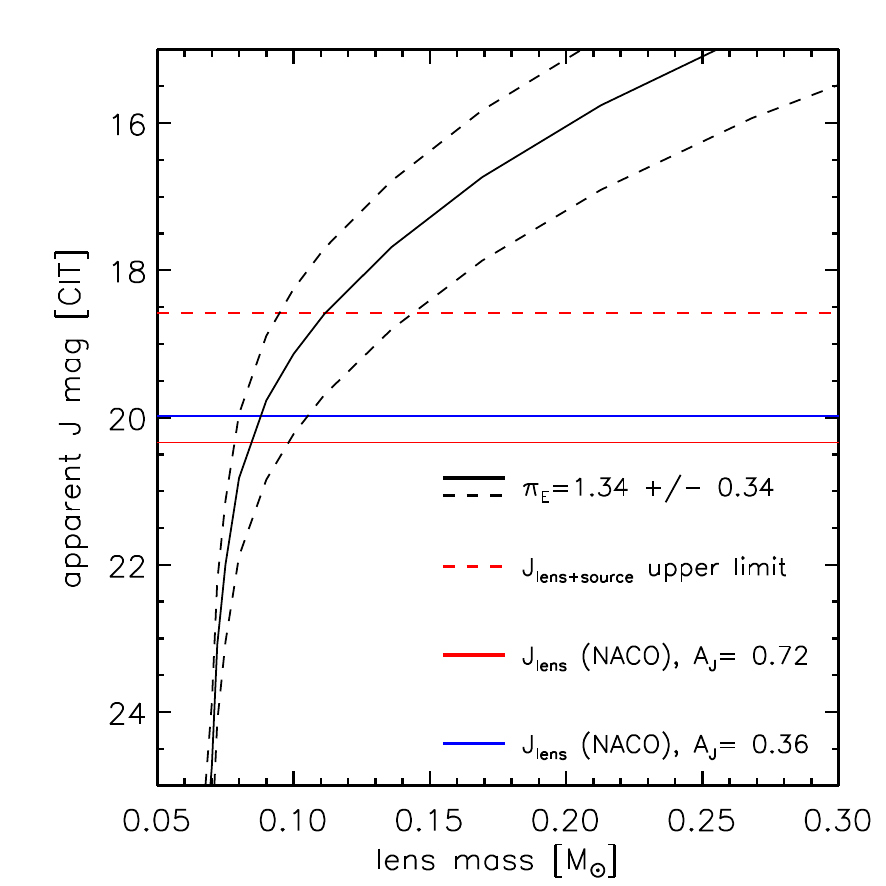}
  \includegraphics[width=7.1cm]{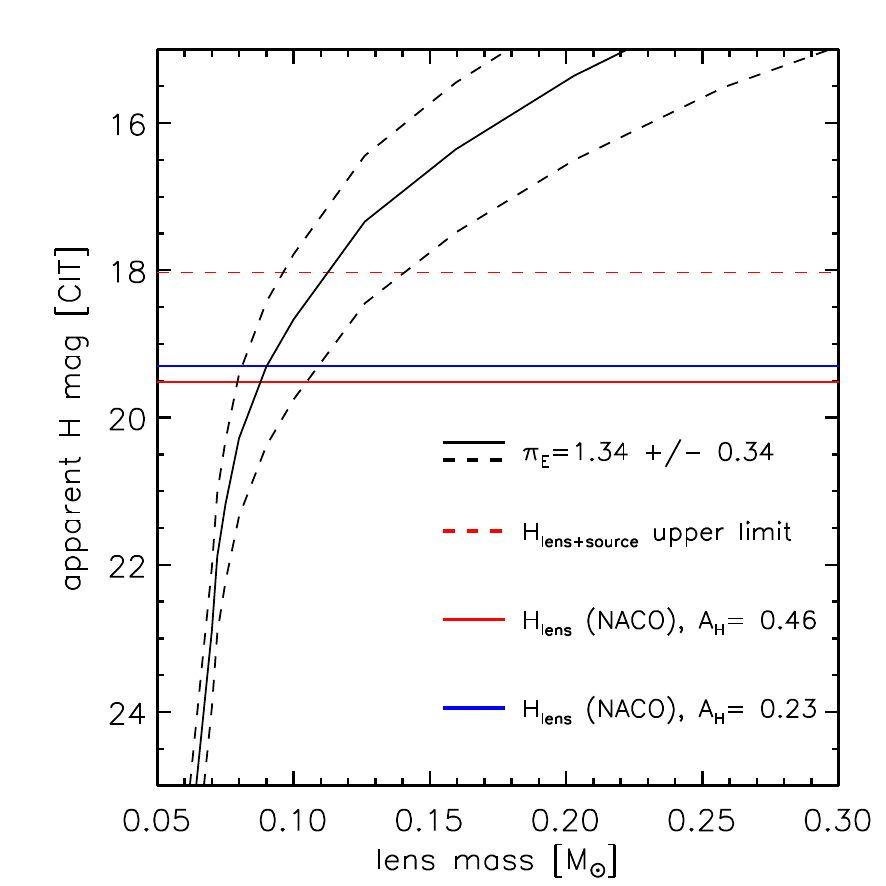}
\caption{Mass-Magnitude relations for K ({\bf top }) , J ({\bf
middle}) and H ({\bf bottom}) bands  [CIT system] derived from
\cite{2000A&A...364..217D} for M-dwarfs (with masses $> 0.10 ~\msun$)
and L-dwarfs (masses $<0.10 \msun$) from
\cite{2000ApJ...542..464C}. The transition between the two relations
at $ \sim 0.10~ \msun$ has been linearly interpolated.  The black
curves show the most likely range of distances for the \planet~system
as found by \cite{2008ApJ...684..663B} and the horizontal lines marks
the estimate for the lens flux from the NACO data as well as the upper
limit of the lens flux from measured lens+source flux for a range of
possible interstellar extinctions.}
         \label{FigLensmassLight}
   \end{figure}

\subsection{Additional constraints from future high angular resolution 
observations}

Another improvement can be achieved by measuring the amplitude and
direction of the relative proper motion of source and lens in
combination with the microlensing modeling of the parallax signal
caused by the Earth's motion. Such physical measurements break a model
degeneracy in the projected Einstein radius $\PRE$
\citep{2007ApJ...660..781B,2008ApJ...684..663B}. In the case of
MOA-2007-BLG-192 the degeneracy is particularly acute because of a
gap in event coverage, {\change with different equally well-fitting models}
requiring widely different projections and hence directions for the
relative proper motion, even though the models yield similar
amplitudes: $\sim 5 ~{\rm mas~yr^{-1}}$.  The measurement of both
$\theta_E$ and $\PRE$ yields the lens system mass $M_L = c^2/(4G)
~\theta_E \PRE$.  Ideally, the relative lens-source proper motion
$\mu_{\rm rel}$ is measured when detecting both the lens star and the
source star as done by \cite{2001Natur.414..617A}.

The two stars will not be fully resolved for many years. However, due
to the unique stability of the HST point spread function (PSF) it is
possible to measure source-lens separations (with position angles)
much smaller than the width of the PSF. This is accomplished by
measuring the elongation of the combined lens-source image due to the
fact it is a combination of two point source images rather than
one. The lens and source stars of MOA-2007-BLG-192 will already be about 25
mas apart in 2012.

Simulations by Bennett et. al. (2007) show that measurements of both
the amplitude and orientation are possible for MOA-2007-BLG-192
already in 2012.  These measurements combined with our modeling will
improve the knowledge of system parameters (masses, orbital
separation) to about 10\%.  The key point is that the direction of the
elongation will give us a measurement of the direction of the relative
lens-source proper motion $\mu_{\rm rel}$ and therefore resolve the
remaining parallax degeneracies {\change
\citep{2004ApJ...615..450G,2007ApJ...660..781B,2008ApJ...684..663B}.}

\subsection{Properties of the planetary system}

The effective temperature of the planet, for the parameters of the
parent star and orbit separation given above, is $47^{+7}_{-8}\rm{K}$
for an albedo of zero, and $40^{+6}_{-7}\rm{K}$ for an albedo of
0.5. Based on observations of a tenuous atmosphere (20-60 microbars)
of nitrogen on Pluto, the temperature of bright surfaces ices on Pluto
at perihelion is estimated to be between 35-40 K
\citep{1999Icar..141..299S}.  Thus, if nitrogen were available, the
surface of this planet might look like that of Pluto on the basis of
stellar heating alone. However, the large mass of the planet compared
with that of Pluto necessitates examining the possible role of heat
from the interior of the planet. The maximum temperature possible with
zero albedo, 54 K, remains below the pure nitrogen melting point of 63
K, and well below the methane melting point of 91 K.

The present-day terrestrial heat flow ($0.087 W/m^2$) value is about
10 times less than the roughly $1 W/m^2$ deposited by the lensing star
on its planet at local noon. Thus the average heat flow coming from
the planet itself will not raise the surface temperature
significantly, even for a fully rocky body three times the mass of the
Earth \citep{2010ttt..work...75L}. Of course, we do not know the age
of the star; were we to use the Hadean heat flow value for the Earth
\citep{2008Natur.456..493H} for the 3.2 Earth mass body, the influx
from geothermal heating could exceed the energy received from the
star. The surface temperature could then be above the nitrogen melting
point, leading to the possibility of liquid nitrogen lakes or seas if
the atmospheric pressure were 0.1 bar or more. The lensing star-planet
system is likely older than this, and hence the planet's heat flow
correspondingly less.

Because the distribution of heat flow on a terrestrial planet can be
strongly heterogeneous, one could imagine places on the surface with
much higher heat flow than the average value for the planet, such that
temperatures might exceed the melting point not just of nitrogen but
of methane. Thus, if sufficient quantities of these molecules were
present, the planet's surface might have zones resembling the
hydrocarbon lakes and seas of Saturn's moon Titan. The possibility of
liquid water cannot be discounted, but would most likely be below the
surface or in very restricted, volcanically active, locales.

\section{Conclusions}
In this study we have presented the analysis of  photometric data
in the near infrared JHKs bands at two different epochs of planetary
microlensing event ~\event~{\bf ,} obtained with the AO system NACO mounted
on UT4 at ESO.  According to the best-fit lens models as given in
\cite{2008ApJ...684..663B} the difference in the magnification of the
source is $0.230 \pm {0.015}$ for the two epochs. If the lens is
non-luminous this would be the expected photometric gradient in our
data set in the absence any blended light contribution. {\change
Our $K$-band data, when combined  with results from previous optical
data, are inconsistent with such a scenario at $4,8 \sigma$}.  In fact
the data imply that there is a significant amount of blended light at
the location of \event. Assuming that this blend is the lens, the data
favor a scenario in which the lens would be a close-by
($660^{+100}_{-70}\rm{pc}$) late M-dwarf.  This is consistent with the
estimates for a stellar lens as based on constraints from extended
source and parallax effects as discussed in
\cite{2008ApJ...684..663B}. While the data available at the time of
the discovery paper were consistent with a broad range of planetary
host masses, the new NACO data presented here support the hypothesis
of a stellar host for \planet. Of course it is conceivable that the
detected blend stems not from the lens, but either from a stellar
companion to the source, the lens or an unrelated background
star. However the probabilities for such scenarios are low and using
Ockham's razor the most likely explanation is that the lens is an
M-dwarf, which implies a planetary mass of $3.2^{+5.2}_{-1.8}\mTer$
for \planet, placing it among the front row of known least massive
cool planets in orbit around one of the least massive host stars.

\planet~ is a landmark exoplanet discovery suggesting that planet
formation occurs down to to the very low mass end of the stellar
population. This is in agreement with the recent statistical results of 
\cite{2012Natur.481..167C} that on average, every star in the Milky Way hosts 
at least one planet. \event~ is the first microlensing event for which multi epoch
AO data has been obtained and demonstrates the usefulness of this
technique for microlensing,  by constraining the physical
characteristics of microlensing planetary systems and providing
important experiences to optimize future AO observations, which
ideally should be carried out in ToO mode for the first epoch, to
ensure the source is still significantly magnified.

\begin{acknowledgements}
We like to say thanks to the ESO Paranal and Garching teams for their
high quality service and support in carrying out the NACO
observations. Especially thanks to ESO staff C.Dumas, C. Lidman and
P. Amico. Thanks also to the IRSF observatory staff.  D.K. and
A.C. especially thank ARI, where part of the work was done during a
visit. Special thanks to the support from ANR via the HOLMES project
grant.  Special thanks to David Warren for supporting the work done at
Canopus Observatory.  D.P.B.\ was supported by grants NNX07AL71G \&
NNX10AI81G from NASA and AST-0708890 from the NSF. Work by S.D. was
performed under contract with the California Institute of Technology
(Caltech) funded by NASA through the Sagan Fellowship Program.  A.G.\
was supported by NSF grant AST-0757888. 
JL work  was financed within the scope of the program 
"Incentivazione alla mobilita' di studiosi straineri e italiani residenti all'estero."
AU was supported from the European Research Council under the European
Community's Seventh Framework Programme (FP7/2007-2013) / ERC grant
agreement no. 246678. D.K. also likes to thank G. James for his graphic support. 
\end{acknowledgements}

%%%%%%%%%%%%%%%%%%%%%%%%%%%%%%%%%%%%%%%%%%%
\bibliographystyle{aa}
\bibliography{ao192}

\begin{thebibliography}{41}
\expandafter\ifx\csname natexlab\endcsname\relax\def\natexlab#1{#1}\fi

\bibitem[{{Alcock} {et~al.}(2001){Alcock}, {Allsman}, {Alves}, {Axelrod},
  {Becker}, {Bennett}, {Cook}, {Drake}, {Freeman}, {Geha}, {Griest}, {Keller},
  {Lehner}, {Marshall}, {Minniti}, {Nelson}, {Peterson}, {Popowski}, {Pratt},
  {Quinn}, {Stubbs}, {Sutherland}, {Tomaney}, {Vandehei}, \&
  {Welch}}]{2001Natur.414..617A}
{Alcock}, C., {Allsman}, R.~A., {Alves}, D.~R., {et~al.} 2001, \nat, 414, 617

\bibitem[{{Beaulieu} {et~al.}(2006){Beaulieu}, {Bennett}, {Fouqu{\'e}},
  {Williams}, {Dominik}, {J{\o}rgensen}, {Kubas}, {Cassan}, {Coutures},
  {Greenhill}, {Hill}, {Menzies}, {Sackett}, {Albrow}, {Brillant}, {Caldwell},
  {Calitz}, {Cook}, {Corrales}, {Desort}, {Dieters}, {Dominis}, {Donatowicz},
  {Hoffman}, {Kane}, {Marquette}, {Martin}, {Meintjes}, {Pollard}, {Sahu},
  {Vinter}, {Wambsganss}, {Woller}, {Horne}, {Steele}, {Bramich}, {Burgdorf},
  {Snodgrass}, {Bode}, {Udalski}, {Szyma{\'n}ski}, {Kubiak}, {Wi{\c e}ckowski},
  {Pietrzy{\'n}ski}, {Soszy{\'n}ski}, {Szewczyk}, {Wyrzykowski},
  {Paczy{\'n}ski}, {Abe}, {Bond}, {Britton}, {Gilmore}, {Hearnshaw}, {Itow},
  {Kamiya}, {Kilmartin}, {Korpela}, {Masuda}, {Matsubara}, {Motomura},
  {Muraki}, {Nakamura}, {Okada}, {Ohnishi}, {Rattenbury}, {Sako}, {Sato},
  {Sasaki}, {Sekiguchi}, {Sullivan}, {Tristram}, {Yock}, \&
  {Yoshioka}}]{2006Natur.439..437B}
{Beaulieu}, J.-P., {Bennett}, D.~P., {Fouqu{\'e}}, P., {et~al.} 2006, \nat,
  439, 437

\bibitem[{{Bennett} {et~al.}(2006){Bennett}, {Anderson}, {Bond}, {Udalski}, \&
  {Gould}}]{2006ApJ...647L.171B}
{Bennett}, D.~P., {Anderson}, J., {Bond}, I.~A., {Udalski}, A., \& {Gould}, A.
  2006, \apjl, 647, L171

\bibitem[{{Bennett} {et~al.}(2007){Bennett}, {Anderson}, \&
  {Gaudi}}]{2007ApJ...660..781B}
{Bennett}, D.~P., {Anderson}, J., \& {Gaudi}, B.~S. 2007, \apj, 660, 781

\bibitem[{{Bennett} {et~al.}(2008){Bennett}, {Bond}, {Udalski}, {Sumi}, {Abe},
  {Fukui}, {Furusawa}, {Hearnshaw}, {Holderness}, {Itow}, {Kamiya}, {Korpela},
  {Kilmartin}, {Lin}, {Ling}, {Masuda}, {Matsubara}, {Miyake}, {Muraki},
  {Nagaya}, {Okumura}, {Ohnishi}, {Perrott}, {Rattenbury}, {Sako}, {Saito},
  {Sato}, {Skuljan}, {Sullivan}, {Sweatman}, {Tristram}, {Yock}, {Kubiak},
  {Szyma{\'n}ski}, {Pietrzy{\'n}ski}, {Soszy{\'n}ski}, {Szewczyk},
  {Wyrzykowski}, {Ulaczyk}, {Batista}, {Beaulieu}, {Brillant}, {Cassan},
  {Fouqu{\'e}}, {Kervella}, {Kubas}, \& {Marquette}}]{2008ApJ...684..663B}
{Bennett}, D.~P., {Bond}, I.~A., {Udalski}, A., {et~al.} 2008, \apj, 684, 663

\bibitem[{{Bennett} \& {Rhie}(1996)}]{1996ApJ...472..660B}
{Bennett}, D.~P. \& {Rhie}, S.~H. 1996, \apj, 472, 660

\bibitem[{{Bennett} {et~al.}(2010){Bennett}, {Rhie}, {Nikolaev}, {Gaudi},
  {Udalski}, {Gould}, {Christie}, {Maoz}, {Dong}, {McCormick}, {Szyma{\'n}ski},
  {Tristram}, {Macintosh}, {Cook}, {Kubiak}, {Pietrzy{\'n}ski},
  {Soszy{\'n}ski}, {Szewczyk}, {Ulaczyk}, {Wyrzykowski}, {The OGLE
  Collaboration}, {DePoy}, {Han}, {Kaspi}, {Lee}, {Mallia}, {Natusch}, {Park},
  {Pogge}, {Polishook}, {The {$\mu$}FUN Collaboration}, {Abe}, {Bond},
  {Botzler}, {Fukui}, {Hearnshaw}, {Itow}, {Kamiya}, {Korpela}, {Kilmartin},
  {Lin}, {Ling}, {Masuda}, {Matsubara}, {Motomura}, {Muraki}, {Nakamura},
  {Okumura}, {Ohnishi}, {Perrott}, {Rattenbury}, {Sako}, {Saito}, {Sato},
  {Skuljan}, {Sullivan}, {Sumi}, {Sweatman}, {Yock}, {The MOA Collaboration},
  {Albrow}, {Allan}, {Beaulieu}, {Bramich}, {Burgdorf}, {Coutures}, {Dominik},
  {Dieters}, {Fouqu{\'e}}, {Greenhill}, {Horne}, {Snodgrass}, {Steele},
  {Tsapras}, {PLANET}, {RoboNet Collaborations}, {Chaboyer}, {Crocker}, \&
  {Frank}}]{2010ApJ...713..837B}
{Bennett}, D.~P., {Rhie}, S.~H., {Nikolaev}, S., {et~al.} 2010, \apj, 713, 837

\bibitem[{{Bessell} \& {Brett}(1988)}]{1988PASP..100.1134B}
{Bessell}, M.~S. \& {Brett}, J.~M. 1988, \pasp, 100, 1134

\bibitem[{{Bond} {et~al.}(2004){Bond}, {Udalski}, {Jaroszy{\'n}ski},
  {Rattenbury}, {Paczy{\'n}ski}, {Soszy{\'n}ski}, {Wyrzykowski},
  {Szyma{\'n}ski}, {Kubiak}, {Szewczyk}, {{\.Z}ebru{\'n}}, {Pietrzy{\'n}ski},
  {Abe}, {Bennett}, {Eguchi}, {Furuta}, {Hearnshaw}, {Kamiya}, {Kilmartin},
  {Kurata}, {Masuda}, {Matsubara}, {Muraki}, {Noda}, {Okajima}, {Sako},
  {Sekiguchi}, {Sullivan}, {Sumi}, {Tristram}, {Yanagisawa}, \&
  {Yock}}]{2004ApJ...606L.155B}
{Bond}, I.~A., {Udalski}, A., {Jaroszy{\'n}ski}, M., {et~al.} 2004, \apjl, 606,
  L155

\bibitem[{{Cassan} {et~al.}(2012){Cassan}, {Kubas}, {Beaulieu}, {Dominik},
  {Horne}, {Greenhill}, {Wambsganss}, {Menzies}, {Williams}, {Jorgensen},
  {Udalski}, {Bennett}, {Albrow}, {Batista}, {Brillant}, {Caldwell}, {Cole},
  {Coutures}, {Cook}, {Dieters}, {Prester}, {Donatowicz}, {Fouque}, {Hill},
  {Kains}, {Kane}, {Marquette}, {Martin}, {Pollard}, {Sahu}, {Vinter},
  {Warren}, {Watson}, {Zub}, {Sumi}, {Szymanski}, {Kubiak}, {Poleski},
  {Soszynski}, {Ulaczyk}, {Pietrzynski}, \&
  {Wyrzykowski}}]{2012Natur.481..167C}
{Cassan}, A., {Kubas}, D., {Beaulieu}, J.-P., {et~al.} 2012, \nat, 481, 167

\bibitem[{{Chabrier} {et~al.}(2000){Chabrier}, {Baraffe}, {Allard}, \&
  {Hauschildt}}]{2000ApJ...542..464C}
{Chabrier}, G., {Baraffe}, I., {Allard}, F., \& {Hauschildt}, P. 2000, \apj,
  542, 464

\bibitem[{{Delfosse} {et~al.}(2000){Delfosse}, {Forveille}, {S{\'e}gransan},
  {Beuzit}, {Udry}, {Perrier}, \& {Mayor}}]{2000A&A...364..217D}
{Delfosse}, X., {Forveille}, T., {S{\'e}gransan}, D., {et~al.} 2000, \aap, 364,
  217

\bibitem[{{Devillard}(1997)}]{1997Msngr..87...19D}
{Devillard}, N. 1997, The Messenger, 87, 19

\bibitem[{{Devillard}(1999)}]{1999ASPC..172..333D}
{Devillard}, N. 1999, in Astronomical Society of the Pacific Conference Series,
  Vol. 172, Astronomical Data Analysis Software and Systems VIII, ed. D.~M.
  {Mehringer}, R.~L. {Plante}, \& D.~A. {Roberts}, 333--+

\bibitem[{{Diolaiti} {et~al.}(2000){Diolaiti}, {Bendinelli}, {Bonaccini},
  {Close}, {Currie}, \& {Parmeggiani}}]{2000A&AS..147..335D}
{Diolaiti}, E., {Bendinelli}, O., {Bonaccini}, D., {et~al.} 2000, \aaps, 147,
  335

\bibitem[{{Dong} {et~al.}(2009){Dong}, {Gould}, {Udalski}, {Anderson},
  {Christie}, {Gaudi}, {The OGLE Collaboration}, {Jaroszy{\'n}ski}, {Kubiak},
  {Szyma{\'n}ski}, {Pietrzy{\'n}ski}, {Soszy{\'n}ski}, {Szewczyk}, {Ulaczyk},
  {Wyrzykowski}, {The {$\mu$}FUN Collaboration}, {DePoy}, {Fox}, {Gal-Yam},
  {Han}, {L{\'e}pine}, {McCormick}, {Ofek}, {Park}, {Pogge}, {The MOA
  Collaboration}, {Abe}, {Bennett}, {Bond}, {Britton}, {Gilmore}, {Hearnshaw},
  {Itow}, {Kamiya}, {Kilmartin}, {Korpela}, {Masuda}, {Matsubara}, {Motomura},
  {Muraki}, {Nakamura}, {Ohnishi}, {Okada}, {Rattenbury}, {Saito}, {Sako},
  {Sasaki}, {Sullivan}, {Sumi}, {Tristram}, {Yanagisawa}, {Yock}, {Yoshoika},
  {The PLANET/Robo Net Collaborations}, {Albrow}, {Beaulieu}, {Brillant},
  {Calitz}, {Cassan}, {Cook}, {Coutures}, {Dieters}, {Prester}, {Donatowicz},
  {Fouqu{\'e}}, {Greenhill}, {Hill}, {Hoffman}, {Horne}, {J{\o}rgensen},
  {Kane}, {Kubas}, {Marquette}, {Martin}, {Meintjes}, {Menzies}, {Pollard},
  {Sahu}, {Vinter}, {Wambsganss}, {Williams}, {Bode}, {Bramich}, {Burgdorf},
  {Snodgrass}, {Steele}, {Doublier}, \& {Foellmi}}]{2009ApJ...695..970D}
{Dong}, S., {Gould}, A., {Udalski}, A., {et~al.} 2009, \apj, 695, 970

\bibitem[{{Duquennoy} \& {Mayor}(1991)}]{1991A&A...248..485D}
{Duquennoy}, A. \& {Mayor}, M. 1991, \aap, 248, 485

\bibitem[{{Gaudi} {et~al.}(2008){Gaudi}, {Bennett}, {Udalski}, {Gould},
  {Christie}, {Maoz}, {Dong}, {McCormick}, {Szyma{\'n}ski}, {Tristram},
  {Nikolaev}, {Paczy{\'n}ski}, {Kubiak}, {Pietrzy{\'n}ski}, {Soszy{\'n}ski},
  {Szewczyk}, {Ulaczyk}, {Wyrzykowski}, {DePoy}, {Han}, {Kaspi}, {Lee},
  {Mallia}, {Natusch}, {Pogge}, {Park}, {Abe}, {Bond}, {Botzler}, {Fukui},
  {Hearnshaw}, {Itow}, {Kamiya}, {Korpela}, {Kilmartin}, {Lin}, {Masuda},
  {Matsubara}, {Motomura}, {Muraki}, {Nakamura}, {Okumura}, {Ohnishi},
  {Rattenbury}, {Sako}, {Saito}, {Sato}, {Skuljan}, {Sullivan}, {Sumi},
  {Sweatman}, {Yock}, {Albrow}, {Allan}, {Beaulieu}, {Burgdorf}, {Cook},
  {Coutures}, {Dominik}, {Dieters}, {Fouqu{\'e}}, {Greenhill}, {Horne},
  {Steele}, {Tsapras}, {Chaboyer}, {Crocker}, {Frank}, \&
  {Macintosh}}]{2008Sci...319..927G}
{Gaudi}, B.~S., {Bennett}, D.~P., {Udalski}, A., {et~al.} 2008, Science, 319,
  927

\bibitem[{{Ghosh} {et~al.}(2004){Ghosh}, {DePoy}, {Gal-Yam}, {Gaudi}, {Gould},
  {Han}, {Lipkin}, {Maoz}, {Ofek}, {Park}, {Pogge}, {Salim}, {Abe}, {Bennett},
  {Bond}, {Eguchi}, {Furuta}, {Hearnshaw}, {Kamiya}, {Kilmartin}, {Kurata},
  {Masuda}, {Matsubara}, {Muraki}, {Noda}, {Okajima}, {Rattenbury}, {Sako},
  {Sekiguchi}, {Sullivan}, {Sumi}, {Tristram}, {Yanagisawa}, {Yock}, {Udalski},
  {Soszy{\'n}ski}, {Wyrzykowski}, {Kubiak}, {Szyma{\'n}ski}, {Pietrzy{\'n}ski},
  {Szewczyk}, {{\.Z}ebru{\'n}}, {Albrow}, {Beaulieu}, {Caldwell}, {Cassan},
  {Coutures}, {Dominik}, {Donatowicz}, {Fouqu{\'e}}, {Greenhill}, {Hill},
  {Horne}, {J{\o}rgensen}, {Kane}, {Kubas}, {Martin}, {Menzies}, {Pollard},
  {Sahu}, {Wambsganss}, {Watson}, \& {Williams}}]{2004ApJ...615..450G}
{Ghosh}, H., {DePoy}, D.~L., {Gal-Yam}, A., {et~al.} 2004, \apj, 615, 450

\bibitem[{{Glass} \& {Nagata}(2000)}]{2000MNSSA..59..110G}
{Glass}, I.~S. \& {Nagata}, T. 2000, Monthly Notes of the Astronomical Society
  of South Africa, 59, 110

\bibitem[{{Gould} {et~al.}(2010){Gould}, {Dong}, {Bennett}, {Bond}, {Udalski},
  \& {Kozlowski}}]{2010ApJ...710.1800G}
{Gould}, A., {Dong}, S., {Bennett}, D.~P., {et~al.} 2010, \apj, 710, 1800

\bibitem[{{Gould} \& {Loeb}(1992)}]{1992ApJ...396..104G}
{Gould}, A. \& {Loeb}, A. 1992, \apj, 396, 104

\bibitem[{{Gould} {et~al.}(2006){Gould}, {Udalski}, {An}, {Bennett}, {Zhou},
  {Dong}, {Rattenbury}, {Gaudi}, {Yock}, {Bond}, {Christie}, {Horne},
  {Anderson}, {Stanek}, {DePoy}, {Han}, {McCormick}, {Park}, {Pogge},
  {Poindexter}, {Soszy{\'n}ski}, {Szyma{\'n}ski}, {Kubiak}, {Pietrzy{\'n}ski},
  {Szewczyk}, {Wyrzykowski}, {Ulaczyk}, {Paczy{\'n}ski}, {Bramich},
  {Snodgrass}, {Steele}, {Burgdorf}, {Bode}, {Botzler}, {Mao}, \&
  {Swaving}}]{2006ApJ...644L..37G}
{Gould}, A., {Udalski}, A., {An}, D., {et~al.} 2006, \apjl, 644, L37

\bibitem[{{Hopkins} {et~al.}(2008){Hopkins}, {Harrison}, \&
  {Manning}}]{2008Natur.456..493H}
{Hopkins}, M., {Harrison}, T.~M., \& {Manning}, C.~E. 2008, \nat, 456, 493

\bibitem[{{Janczak} {et~al.}(2010){Janczak}, {Fukui}, {Dong}, {Monard},
  {Koz{\l}owski}, {Gould}, {Beaulieu}, {Kubas}, {Marquette}, {Sumi}, {Bond},
  {Bennett}, {Abe}, {Furusawa}, {Hearnshaw}, {Hosaka}, {Itow}, {Kamiya},
  {Korpela}, {Kilmartin}, {Lin}, {Ling}, {Makita}, {Masuda}, {Matsubara},
  {Miyake}, {Muraki}, {Nagaya}, {Nagayama}, {Nishimoto}, {Ohnishi}, {Perrott},
  {Rattenbury}, {Sako}, {Saito}, {Skuljan}, {Sullivan}, {Sweatman}, {Tristram},
  {Yock}, {The MOA Collaboration}, {An}, {Christie}, {Chung}, {DePoy}, {Gaudi},
  {Han}, {Lee}, {Mallia}, {Natusch}, {Park}, {Pogge}, {The {$\mu$}FUN
  Collaboration}, {Anguita}, {Calchi Novati}, {Dominik}, {J{\o}rgensen},
  {Masi}, {Mathiasen}, {The MiNDSTEp Collaboration}, {Batista}, {Brillant},
  {Cassan}, {Cole}, {Corrales}, {Coutures}, {Dieters}, {Fouqu{\'e}},
  {Greenhill}, \& {The PLANET Collaboration}}]{2010ApJ...711..731J}
{Janczak}, J., {Fukui}, A., {Dong}, S., {et~al.} 2010, \apj, 711, 731

\bibitem[{{Kato} {et~al.}(2007){Kato}, {Nagashima}, {Nagayama}, {Kurita},
  {Koerwer}, {Kawai}, {Yamamuro}, {Zenno}, {Nishiyama}, {Baba}, {Kadowaki},
  {Haba}, {Hatano}, {Shimizu}, {Nishimura}, {Nagata}, {Sato}, {Murai},
  {Kawazu}, {Nakajima}, {Nakaya}, {Kandori}, {Kusakabe}, {Ishihara},
  {Kaneyasu}, {Hashimoto}, {Tamura}, {Tanab{\'e}}, {Ita}, {Matsunaga},
  {Nakada}, {Sugitani}, {Wakamatsu}, {Glass}, {Feast}, {Menzies}, {Whitelock},
  {Fourie}, {Stoffels}, {Evans}, \& {Hasegawa}}]{2007PASJ...59..615K}
{Kato}, D., {Nagashima}, C., {Nagayama}, T., {et~al.} 2007, \pasj, 59, 615

\bibitem[{{Kubas} {et~al.}(2008){Kubas}, {Cassan}, {Dominik}, {Bennett},
  {Wambsganss}, {Brillant}, {Beaulieu}, {Albrow}, {Batista}, {Bode}, {Bramich},
  {Burgdorf}, {Caldwell}, {Calitz}, {Cook}, {Coutures}, {Dieters}, {Dominis
  Prester}, {Donatowicz}, {Fouqu{\'e}}, {Greenhill}, {Hill}, {Hoffman},
  {Horne}, {J{\o}rgensen}, {Kains}, {Kane}, {Marquette}, {Martin}, {Meintjes},
  {Menzies}, {Pollard}, {Sahu}, {Snodgrass}, {Steele}, {Tsapras}, {Vinter},
  {Williams}, {Woller}, {Zub}, \& {The PLANET/Robonet
  Collaboration}}]{2008A&A...483..317K}
{Kubas}, D., {Cassan}, A., {Dominik}, M., {et~al.} 2008, \aap, 483, 317

\bibitem[{{Leggett} {et~al.}(2010){Leggett}, {Burningham}, {Saumon}, {Marley},
  {Warren}, {Smart}, {Jones}, {Lucas}, {Pinfield}, \&
  {Tamura}}]{2010ApJ...710.1627L}
{Leggett}, S.~K., {Burningham}, B., {Saumon}, D., {et~al.} 2010, \apj, 710,
  1627

\bibitem[{{Lenzen} {et~al.}(2003){Lenzen}, {Hartung}, {Brandner}, {Finger},
  {Hubin}, {Lacombe}, {Lagrange}, {Lehnert}, {Moorwood}, \&
  {Mouillet}}]{2003SPIE.4841..944L}
{Lenzen}, R., {Hartung}, M., {Brandner}, W., {et~al.} 2003, in Society of
  Photo-Optical Instrumentation Engineers (SPIE) Conference S eries, Vol. 4841,
  Society of Photo-Optical Instrumentation Engineers (SPIE) Conference S eries,
  ed. {M.~Iye \& A.~F.~M.~Moorwood}, 944--952

\bibitem[{{Liebes}(1964)}]{1964PhRv..133..835L}
{Liebes}, S. 1964, Physical Review, 133, 835

\bibitem[{{Lunine}(2010)}]{2010ttt..work...75L}
{Lunine}, J. 2010, in Through Time; A Workshop On Titan's Past, Present and
  Future, ed. {V.~Cottini, C.~Nixon, \& R.~Lorenz}, 75--+

\bibitem[{{Mao} \& {Paczynski}(1991)}]{1991ApJ...374L..37M}
{Mao}, S. \& {Paczynski}, B. 1991, \apjl, 374, L37

\bibitem[{{Marigo} {et~al.}(2008){Marigo}, {Girardi}, {Bressan}, {Groenewegen},
  {Silva}, \& {Granato}}]{2008A&A...482..883M}
{Marigo}, P., {Girardi}, L., {Bressan}, A., {et~al.} 2008, \aap, 482, 883

\bibitem[{{Marshall} {et~al.}(2006){Marshall}, {Robin}, {Reyl{\'e}},
  {Schultheis}, \& {Picaud}}]{2006A&A...453..635M}
{Marshall}, D.~J., {Robin}, A.~C., {Reyl{\'e}}, C., {Schultheis}, M., \&
  {Picaud}, S. 2006, \aap, 453, 635

\bibitem[{{Nagayama} {et~al.}(2003){Nagayama}, {Nagashima}, {Nakajima},
  {Nagata}, {Sato}, {Nakaya}, {Yamamuro}, {Sugitani}, \&
  {Tamura}}]{2003SPIE.4841..459N}
{Nagayama}, T., {Nagashima}, C., {Nakajima}, Y., {et~al.} 2003, in Society of
  Photo-Optical Instrumentation Engineers (SPIE) Conference Series, Vol. 4841,
  Society of Photo-Optical Instrumentation Engineers (SPIE) Conference Series,
  ed. {M.~Iye \& A.~F.~M.~Moorwood}, 459--464

\bibitem[{{Rousset} {et~al.}(2003){Rousset}, {Lacombe}, {Puget}, {Hubin},
  {Gendron}, {Fusco}, {Arsenault}, {Charton}, {Feautrier}, {Gigan}, {Kern},
  {Lagrange}, {Madec}, {Mouillet}, {Rabaud}, {Rabou}, {Stadler}, \&
  {Zins}}]{2003SPIE.4839..140R}
{Rousset}, G., {Lacombe}, F., {Puget}, P., {et~al.} 2003, in Society of
  Photo-Optical Instrumentation Engineers (SPIE) Conference Series, Vol. 4839,
  Society of Photo-Optical Instrumentation Engineers (SPIE) Conference Series,
  ed. {P.~L.~Wizinowich \& D.~Bonaccini}, 140--149

\bibitem[{{Schechter} {et~al.}(1993){Schechter}, {Mateo}, \&
  {Saha}}]{1993PASP..105.1342S}
{Schechter}, P.~L., {Mateo}, M., \& {Saha}, A. 1993, \pasp, 105, 1342

\bibitem[{{Schlegel} {et~al.}(1998){Schlegel}, {Finkbeiner}, \&
  {Davis}}]{1998ApJ...500..525S}
{Schlegel}, D.~J., {Finkbeiner}, D.~P., \& {Davis}, M. 1998, \apj, 500, 525

\bibitem[{{Stansberry} \& {Yelle}(1999)}]{1999Icar..141..299S}
{Stansberry}, J.~A. \& {Yelle}, R.~V. 1999, \icarus, 141, 299

\bibitem[{{Sumi} {et~al.}(2010){Sumi}, {Bennett}, {Bond}, {Udalski}, {Batista},
  {Dominik}, {Fouqu{\'e}}, {Kubas}, {Gould}, {Macintosh}, {Cook}, {Dong},
  {Skuljan}, {Cassan}, {Abe}, {Botzler}, {Fukui}, {Furusawa}, {Hearnshaw},
  {Itow}, {Kamiya}, {Kilmartin}, {Korpela}, {Lin}, {Ling}, {Masuda},
  {Matsubara}, {Miyake}, {Muraki}, {Nagaya}, {Nagayama}, {Ohnishi}, {Okumura},
  {Perrott}, {Rattenbury}, {Saito}, {Sako}, {Sullivan}, {Sweatman}, {Tristram},
  {Yock}, {The MOA Collaboration}, {Beaulieu}, {Cole}, {Coutures}, {Duran},
  {Greenhill}, {Jablonski}, {Marboeuf}, {Martioli}, {Pedretti}, {Pejcha},
  {Rojo}, {Albrow}, {Brillant}, {Bode}, {Bramich}, {Burgdorf}, {Caldwell},
  {Calitz}, {Corrales}, {Dieters}, {Dominis Prester}, {Donatowicz}, {Hill},
  {Hoffman}, {Horne}, {J{\o}rgensen}, {Kains}, {Kane}, {Marquette}, {Martin},
  {Meintjes}, {Menzies}, {Pollard}, {Sahu}, {Snodgrass}, {Steele}, {Street},
  {Tsapras}, {Wambsganss}, {Williams}, {Zub}, {The PLANET Collaboration},
  {Szyma{\'n}ski}, {Kubiak}, {Pietrzy{\'n}ski}, {Soszy{\'n}ski}, {Szewczyk},
  {Wyrzykowski}, {Ulaczyk}, {The OGLE Collaboration}, {Allen}, {Christie},
  {DePoy}, {Gaudi}, {Han}, {Janczak}, {Lee}, {McCormick}, {Mallia}, {Monard},
  {Natusch}, {Park}, {Pogge}, {Santallo}, \& {The {$\mu$}FUN
  Collaboration}}]{2010ApJ...710.1641S}
{Sumi}, T., {Bennett}, D.~P., {Bond}, I.~A., {et~al.} 2010, \apj, 710, 1641

\bibitem[{{Udalski} {et~al.}(2005){Udalski}, {Jaroszy{\'n}ski},
  {Paczy{\'n}ski}, {Kubiak}, {Szyma{\'n}ski}, {Soszy{\'n}ski},
  {Pietrzy{\'n}ski}, {Ulaczyk}, {Szewczyk}, {Wyrzykowski}, {Christie}, {DePoy},
  {Dong}, {Gal-Yam}, {Gaudi}, {Gould}, {Han}, {L{\'e}pine}, {McCormick},
  {Park}, {Pogge}, {Bennett}, {Bond}, {Muraki}, {Tristram}, {Yock}, {Beaulieu},
  {Bramich}, {Dieters}, {Greenhill}, {Hill}, {Horne}, \&
  {Kubas}}]{2005ApJ...628L.109U}
{Udalski}, A., {Jaroszy{\'n}ski}, M., {Paczy{\'n}ski}, B., {et~al.} 2005,
  \apjl, 628, L109

\end{thebibliography}
%%%%%%%%%%%%%%%%%%%%%%%%%%%%%%%%%%%%%%%

\appendix

\section{Zeropoints and Uncertainties}

%%%%%%%%%%%%%%%%%%%%%%%%%%

\begin{table*}\label{ZPtable}

\caption{Overview of applied calibrations and transformations between
the photometric instrumental systems of IRSF and NACO into the 2MASS
system. We note that the derived zeropoints for NACO are consistent
with zeropoints from NACO based on regularly taken standard stars
(after correction for atmospheric extinction) and that we do not find
a color term between the NACO and 2MASS system.}  \centering
\begin{tabular}{llllcl} 
\hline\hline             
Band & photometric calibration & number of stars used (after last clipping) & & \\
\hline
  \multicolumn{6}{c}{\it IRSF Single Epoch}\\ % To combine 6 columns into a single one
\hline
  J & $J_{\rm IRSF,2MASS} = 22.854 \pm 0.005 + J_{\rm IRSF,inst} - 0.046 (J_{\rm IRSF,inst}-H_{\rm IRSF,inst}) +  0.015$ &  279&&   \\ 
  H &$H_{\rm IRSF,2MASS} = 22.919 \pm 0.003 + H_{\rm   IRSF,inst} + 0.016 (J_{\rm  IRSF,inst}-H_{\rm  IRSF,inst}) + 0.024$  & 451 && \\
  Ks &  $K_{\rm IRSF,2MASS} = 22.146 \pm 0.003 + K_{\rm  IRSF,inst} + 0.010 (J_{\rm  IRSF,inst}-K_{\rm  IRSF,inst}) + 0.014$& 502 && \\
\hline\\
  \multicolumn{6}{c}{\it NACO Epoch 1}\\ % To combine 6 columns into a single one
    \multicolumn{6}{l}{\it ~Zeropoints calibrated against IRSF    } \\
    \hline
\multicolumn{6}{l}{\it  ~using star "1" (adopted) $~~~~~~~~~~~~~~~~~~~~~~~~~~~~~~~~~~~~~~~~~~~~~~~~~~~~~~~~~~~~~~~~~~~~~~~~~~~~~~~~~~~~~~~~~~~~~~~~~~~~~$ stars "1"+"2"  } \\

\hline
  J & $24.247\pm 0.018 +J_{\rm NACO,inst}$  &($24.254\pm0.007$)   &&  \\ 
  H &  $24.012\pm 0.017 +H_{\rm NACO,inst}$ &($23.987\pm0.028$) && \\
  Ks &  $23.128 \pm 0.017+K_{\rm NACO,inst}$  & ($23.105\pm 0.020$)& &  \\
  
\hline\\
  \multicolumn{6}{c}{\it NACO Epoch 2} \\
    \multicolumn{6}{l}{\it ~Zeropoints calibrated against IRSF    } \\
    \hline
\multicolumn{6}{l}{\it  ~using star "1" (adopted) $~~~~~~~~~~~~~~~~~~~~~~~~~~~~~~~~~~~~~~~~~~~~~~~~~~~~~~~~~~~~~~~~~~~~~~~~~~~~~~~~~~~~~~~~~~~~~~~~~~~~~$ stars "1"+"2"  } \\
\hline
  J & $24.315\pm 0.018 +J_{\rm NACO,inst}$  &($24.345\pm0.030$)   &&  \\ 
  H &  $24.024\pm 0.017 +H_{\rm NACO,inst}$ &($24.030\pm0.060$) && \\
  Ks &  $23.067 \pm 0.017+K_{\rm NACO,inst}$  & ($23.091\pm 0.023$)& &  \\
  
\hline\\

   \multicolumn{6}{l}{\it ~Zeropoints aligned with respect to Epoch 1 [within 4" of target]} \\
   \hline
 J & $24.289 \pm 0.019  +J_{\rm NACO,inst}$&8  &  &  & \\
  H & $24.036\pm  0.012+H_{\rm NACO,inst}$&10 &       & & \\
  Ks & $ 23.116\pm 0.008+K_{\rm NACO,inst}$ &10&  &     & \\
  \hline

\hline
\end{tabular}
\end{table*}

\begin{table*}\label{tab-mcmc}

\caption{Parameter Values and MCMC Uncertainties} 
\centering

\begin{tabular}{ccc} 
parameter & value & 2-$\sigma$ range \\
\hline

$M$  & $0.084{+0.015\atop -0.012} \msun$ & 0.062--$0.120\msun$ \\[0.15cm]
$m$  & $3.2{+5.2\atop -1.8} \mearth$ & 0.8--$14.8\mearth$ \\[0.15cm]
$a$  & $0.66{+0.51\atop -0.22} \,$AU & 0.35--$3.17\,$AU \\[0.15cm]
$D_L$ & $0.70{+0.21\atop -0.12} \,$kpc & 0.5--$1.4\,$kpc \\[0.15cm]
$I_S$ &  $21.59\pm 0.07$ & 21.46--21.64 \\[0.15cm]
$q$ & $1.1 {+1.9\atop -0.6} \times 10^{-4}$ & 0.3--$5.2\times 10^{-4}$ \\[0.15cm]
\hline
\end{tabular}
\end{table*}

\begin{table*}
\caption{This table shows the fit parameters for the 8 distinct
planetary models for MOA-2007-BLG-192 consistent with the NACO data.
$t_0^\prime = t_0 -4240\,$days. $t_0$ and $u_0$ are the time and
distance of the closest approach of the source to the lens
center-of-mass.  $q$ and $d$ are the planet:star mass ratio and
separation, and $\theta$ is the angle between the source trajectory
and the planet-star axis. $I_s$ is the best-fit source magnitude, and
$\pi_E$ and $\phi_E$ are the magnitude and angle of the microlensing
parallax vector. The units for the Einstein radius crossing time,
$t_E$, the source radius crossing time, $t_\ast$, and $t_0^\prime$ are
days, and all other parameters are dimensionless.}
\label{fittable}
\centering
\begin{tabular}{cccccccccccc} 
\hline
Fit  & $\chi^2$ & $t_E$ & $t_0^\prime$ & $u_0$ &  $d$ & $\theta$ & $q$ &
 $t_\ast$ & $I_s$ & $\pi_E$ & $\phi_E$ \\
\hline      
A  & 1121.12 & 82.5 & 5.442 & 0.00309 & 0.912 & $236.9^\circ$ & $3.7\times 10^{-5}$ & 0.040 & 
        21.61 & 1.51 & $208.5^\circ$ \\
B  & 1121.16 & 83.2 & 5.442 & 0.00306 & 1.120 & $236.8^\circ$ & $3.7\times 10^{-5}$ & 0.041 & 
        21.62 & 1.49 & $208.7^\circ$ \\
C  & 1121.66 & 82.9 & 5.456 & -0.00349 & 0.807 & $105.5^\circ$ & $3.4\times 10^{-4}$ & 0.041 & 
        21.60 & 1.47 & $209.2^\circ$ \\
D  & 1122.08 & 83.5 & 5.448 & -0.00296 & 1.113 & $121.9^\circ$ & $5.8\times 10^{-5}$ & 0.041 & 
        21.62 & 1.44 & $209.8^\circ$ \\
E  & 1125.41 & 83.2 & 5.454 & 0.00306 & 0.890 & $240.1^\circ$ & $7.6\times 10^{-5}$ & 0.048 & 
        21.61 & 1.19 & $337.5^\circ$ \\
F  & 1125.44 & 82.5 & 5.454 & 0.00309 & 1.118 & $239.9^\circ$ & $7.5\times 10^{-5}$ & 0.049 & 
        21.60 & 1.16 & $332.3^\circ$ \\
G  & 1125.48 & 81.8 & 5.450 & -0.00313 & 0.897 & $120.3^\circ$ & $6.4\times 10^{-5}$ & 0.049 & 
        21.60 & 1.13 & $336.5^\circ$ \\
H  & 1125.50 & 80.8 & 5.450 & -0.00309 & 1.110 & $120.6^\circ$ & $6.1\times 10^{-5}$ & 0.048 & 
        21.61 & 1.19 & $337.8^\circ$ \\
\hline
\end{tabular}

\end{table*}\label{tabsurvive}

\end{document}